\documentclass[reprint,prb,twocolumn,longbibliography,noeprint,superscriptaddress,floatfix]{revtex4-2}
\usepackage{float}
\usepackage{physics}
\usepackage{graphicx}
\usepackage{xcolor}
\usepackage{amsmath,,amsfonts,amssymb}
\usepackage{textcomp}
\usepackage{bbm}
\usepackage{dsfont}
\usepackage{microtype}
\usepackage{mathtools}
\usepackage{multirow}
\usepackage[utf8]{inputenc}
\usepackage[T1]{fontenc}

\usepackage[pdftex,bookmarks=true,bookmarksopen,bookmarksnumbered,
                colorlinks,
                linkcolor=blue,
                citecolor=blue,
                colorlinks = true,
                urlcolor  = blue,
                anchorcolor = blue
                ]{hyperref}
\usepackage{cleveref}
\usepackage{lineno}
\usepackage{accents}
\usepackage{comment}
\usepackage{enumitem}
\usepackage{textcomp}
\usepackage{setspace}

\begin{document}

\title{Jellybean quantum dots in silicon for qubit coupling and on-chip quantum chemistry}

\author{Zeheng Wang}
\email[]{zenwang@outlook.com}
\author{MengKe Feng}
\author{Santiago Serrano}
\author{William Gilbert}
\author{Ross C. C. Leon}
\author{Tuomo Tanttu}
\author{Philip Mai}
\author{Dylan Liang}
\author{Jonathan Y. Huang}
\author{Yue Su}
\author{Wee Han Lim}
\author{Fay E. Hudson}
\author{Christopher C. Escott}
\author{Andrea Morello}
\author{Chih Hwan Yang}
\author{Andrew S. Dzurak}
\author{Andre Saraiva}
\author{Arne Laucht*}
\email[]{a.laucht@unsw.edu.au}
\affiliation{
 School of Electrical Engineering and Telecommunications,
 The University of New South Wales, Sydney, NSW 2052, Australia
}


\begin{abstract}
The small size and excellent integrability of silicon metal-oxide-semiconductor (SiMOS) quantum dot spin qubits make them an attractive system for mass-manufacturable, scaled-up quantum processors.
Furthermore, classical control electronics can be integrated on-chip, in-between the qubits, if an architecture with sparse arrays of qubits is chosen. 
In such an architecture qubits are either transported across the chip via shuttling, or coupled via mediating quantum systems over short-to-intermediate distances.
This paper investigates the charge and spin characteristics of an elongated quantum dot -- a so-called jellybean quantum dot --  for the prospects of acting as a qubit-qubit coupler. 
Charge transport, charge sensing and magneto-spectroscopy measurements are performed on a SiMOS quantum dot device at mK temperature, and compared to Hartree-Fock multi-electron simulations. 
At low electron occupancies where disorder effects and strong electron-electron interaction dominate over the electrostatic confinement potential, the data reveals the formation of three coupled dots, akin to a tunable, artificial molecule. One dot is formed centrally under the gate and two are formed at the edges.
At high electron occupancies, these dots merge into one large dot with well-defined spin states, verifying that jellybean dots have the potential to be used as qubit couplers in future quantum computing architectures.
\end{abstract}

\maketitle

\section{Introduction}
Semiconductor-based quantum computing architectures have been extensively studied for decades because of their high potential to fulfil the five DiVincenzo criteria for realizing a scalable quantum computer~\cite{divincenzo2000physical,zhang2018qubits,laucht2021roadmap,chatterjee2021semiconductor}. 
Notably, silicon-based, complementary MOS (CMOS)-compatible approaches have been considered especially promising in the race to the quantum advantage and beyond, thanks to the strong scaling and integration capabilities enabled by the mature CMOS industry~\cite{gonzalez-zalba2021scaling}.
Furthermore, benefiting from isotopically enriched $^{28}$Si substrates~\cite{itoh2014isotope,saraiva2022materials} and advanced device architectures, long coherence times, high-speed operation, and high-fidelity universal one- and two-qubit logic gates have been successfully demonstrated in Si-based quantum bits (qubits)~\cite{muhonen2014storing,veldhorst2014addressable,takeda2016faulttolerant,yoneda2018quantumdot,huang2019fidelity,xue2019benchmarking,madzik2022precision,xue2022quantum,noiri2022fast,mills2022high}. 

In addition to the five DiVincenzo criteria, large-scale quantum chip integration requires that the quantum information can a) be transferred between the computing qubit and the transferring qubit, and b) be conserved during the transportation~\cite{divincenzo2000physical}. To this end, some prospective strategies for semiconductor-based quantum devices have already been proposed in theory~\cite{mehl2014twoqubit,srinivasa2015tunable,stano2015fast,mohiyaddin2016transport,vandersypen2017interfacing}. As preliminary attempts to implement these proposals, multi-dot interaction or short distance spin-transfer has been demonstrated in both GaAs~\cite{baart2016singlespin,fujita2017coherent,fedele2021simultaneous} and Si-based devices~\cite{sigillito2019siteselective,yoneda2021coherent,noiri2022shuttlingbased}. 
A promising alternative to having chains of dots is to integrate a single, elongated dot, known as a jellybean dot, which enables coherent spin transport between two quantum dots located some distance apart~\cite{croot2018device,malinowski2018spin,malinowski2019fast}. Since a single jellybean dot can potentially replace a chain of multiple dots, the complexity of the device architecture is reduced and thus favors scalability. 

Maybe not surprisingly, within a jellybean dot, the electron wavefunction is more spread out than in a typical, more symmetric dot, which means that disorder effects and electron-electron interactions may dominate the electron distribution, possibly even leading to the formation of a Wigner molecule~\cite{yannouleas1999spontaneous,cavaliere2009transport,corrigan2021coherent,abadillo-uriel2021twobody,yannouleas2022molecular,yannouleas2022wigner,jang2022wignermolecularizationenabled}. These effects significantly impact the charge characteristics and spin-transfer mechanism of the elongated dot. In addition to providing deep insights into the mesoscopic physics of solid state systems, the formation of a Wigner molecule within a jellybean dot could also be utilized to effectively simulate the quantum dynamics of a molecule or even a polymer~\cite{kiczynski2022engineering}. As such, the characteristics and properties of jellybean dots need to be understood and experimentally studied.

\begin{figure*}[hbt!]
\centering
  \includegraphics[width=\linewidth]{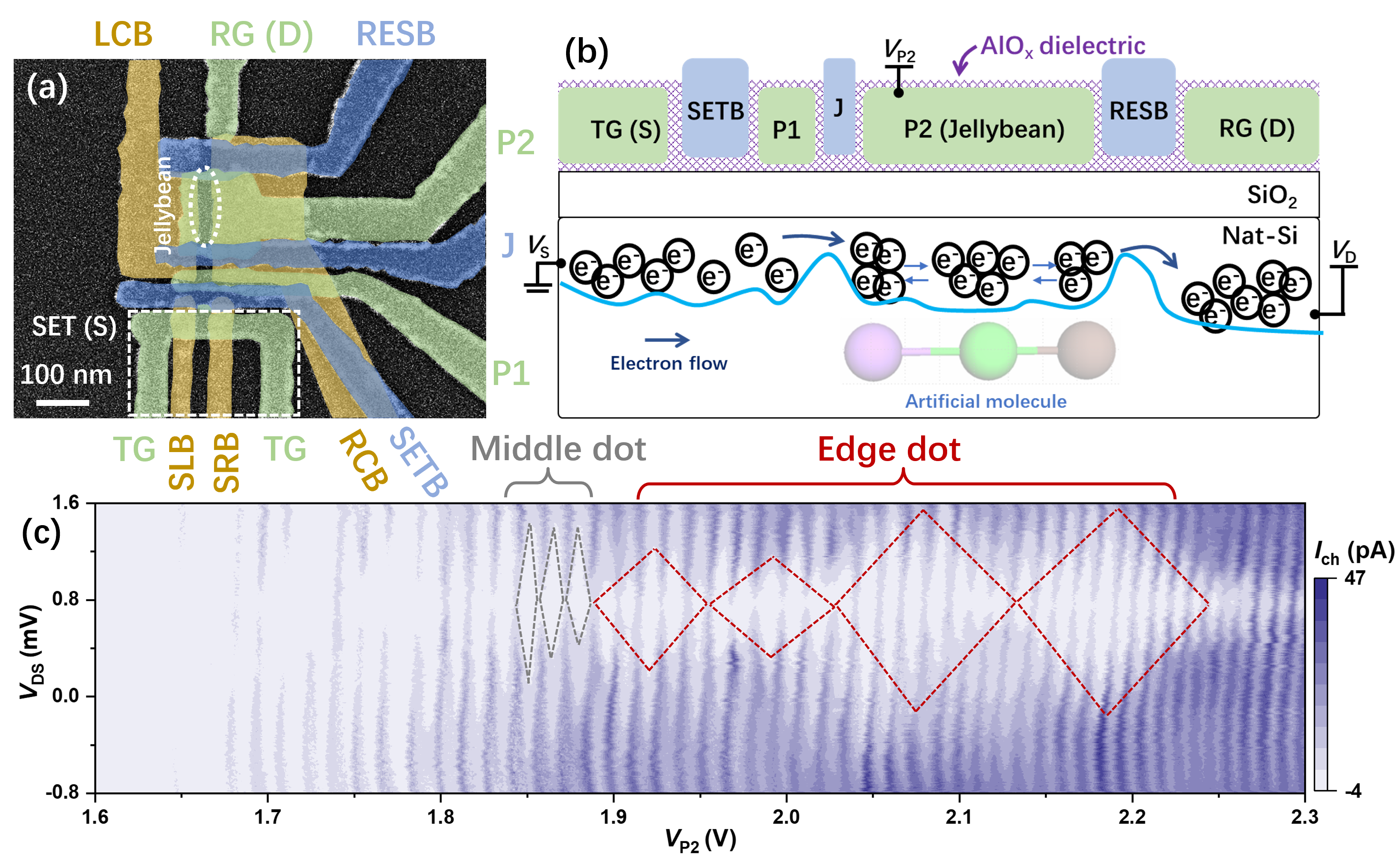}
  \caption{
  Device architecture and jellybean transport measurement.
  (a) False-colored scanning electron microscope (SEM) image of a nominally identical device with gate labels. The location of the jellybean dot is indicated by the dotted ellipse.
  (b) Schematic cross-section of the device along the dot channel, and dot filling in transport mode.
  (c) Coulomb diamond measurement of P2 dot, where two types of the dots with different capacitive coupling to P2 are outlined. 
  }
  \label{fig:device}
\end{figure*}

In this paper, we investigate the charge characteristics of a jellybean dot fabricated on a silicon chip. We find that the charging pattern of our jellybean dot differs from that of a typical round dot in that it initially tends to form three distinct smaller dots (artificial atoms) that eventually merge into a larger one. The formation of this artificial molecule may be explained with either disorder effects or electron-electron interactions. Experimentally, the signatures of these two effects are very difficult to separate, and we perform theoretical calculations to understand the different types of electronic states formed at different dot occupations. The experiments show that we can tune the distance between two artificial atoms of this artificial molecule using a side barrier gate (J). Using field-effect model simulations, we confirm that the side barrier gate J can be used to tune the distance between the two leftmost artificial atoms by up to 33~nm, more than 40\% of the minimum separation. Finally, magneto-spectroscopy measurements show that for low electron occupancies, the total spin behavior does not obey the shell filling pattern expected for a round dot, however after the atoms of the artificial molecule merge into a single dot, the spin filling follows a distinguishable pattern. These results indicate that the jellybean dot is a promising platform to understand mesoscopic physics, and to perform spin-based quantum operations such as Pauli spin blockade, electron-spin resonance and coherent, long-distance spin transport in a Si-based quantum chip.

\section{Charge Characterization of a Jellybean Dot}

We start by characterizing the electrical properties of the device. Figure~\ref{fig:device}(a) shows a false-colored scanning electron microscope (SEM) image of a device nominally identical to the one measured here. The devices are fabricated on a natural silicon substrate with a stack of palladium gate electrodes that are separated by atomic-layer-deposition grown aluminium oxide~\cite{zhao2019singlespin}. The different colors in Fig.~\ref{fig:device}(a) indicate the separate fabrication layers (1st layer: gold; 2nd layer: green; 3rd layer: blue). Gate P2 controls the electrostatic potential of the jellybean dot (width: $\approx30$~nm; length: $\approx150$~nm), while gate RESB controls the tunnel rates between the jellybean and an electron reservoir formed by an ohmic contact (D) which runs underneath the metallic gate RG. A Single-Electron Transistor (SET) can be used as a charge sensor or act as a second electron reservoir for transport measurements by accumulating the electrons from one of the SET ohmics (S) underneath the metallic gate TG. The tunnel rate between the jellybean dot and the second reservoir is controlled by gate J. Gates LCB and RCB provide lateral confinement of the quantum dot channel.

\begin{figure*}[ht!]
\centering
  \includegraphics[width=\linewidth]{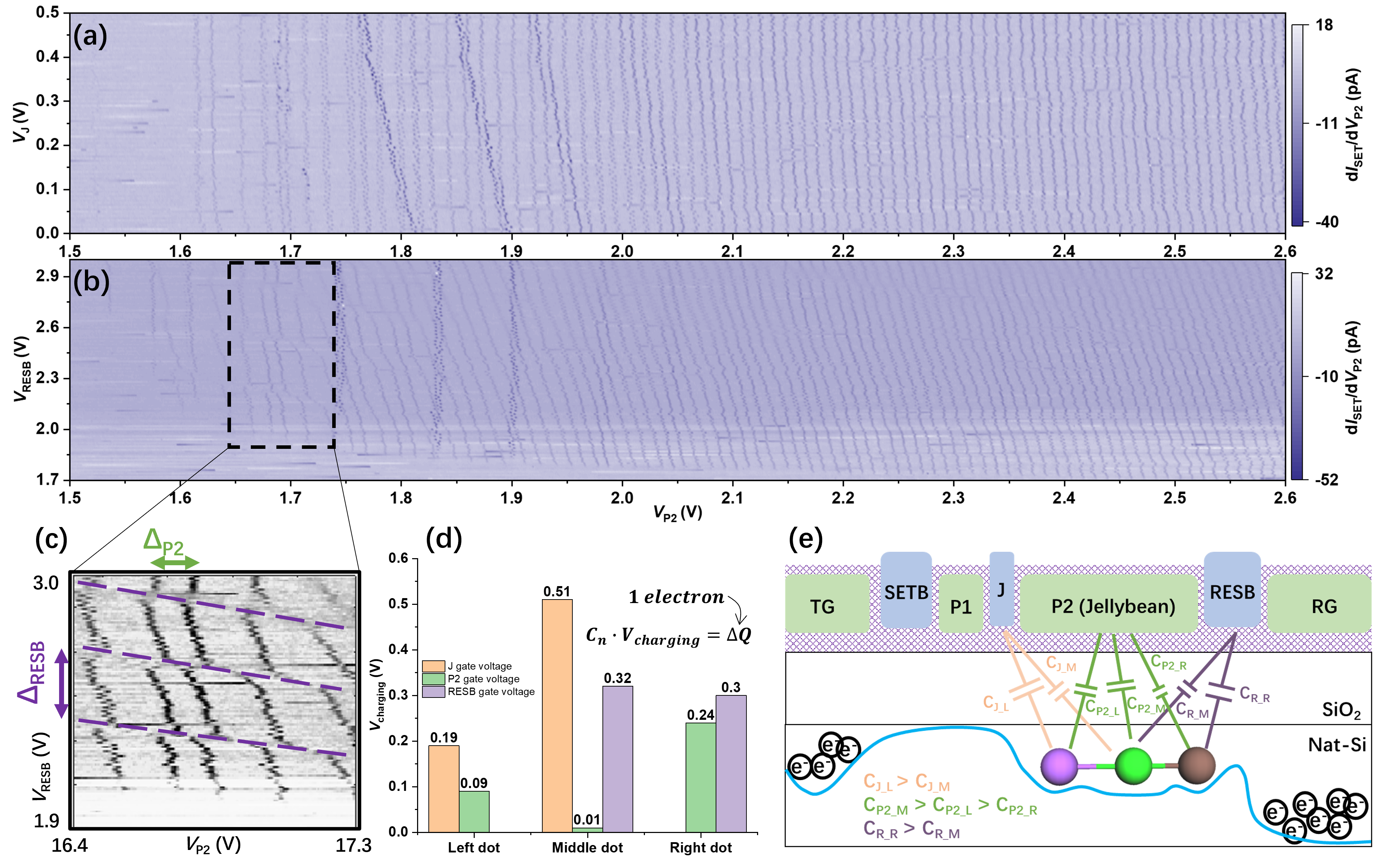}
  \caption{
  Charge occupations measurements using SET charge sensing.
  (a),(b) Charge transitions controlled by P2 together with (a) the J gate and (b) the RESB gate. 
  (c) Zoomed-in structure of the transitions in (b).
  (d) Charging voltage of each of the dots with respect to each of the gates.
  (e) Schematic sketch of the capacitive couplings and their strengths of the dots under P2.
  }
  \label{fig:stability}
\end{figure*}

Figure~\ref{fig:device}(b) shows a schematic cross-section of the device and an illustration of the distribution of electrons and conduction band energies during the transport measurement. The gates SETB and P1 were fully turned-on ($V_{\rm SETB}=V_{\rm P1}=V_{\rm RG}=3$~V to extend the electron reservoir towards the P2 dot. The J and RESB gates were tuned to the sub-threshold regime ($V_{\rm J}=2.6$~V, $V_{\rm RESB}=2.7$~V) to create the barriers confining the dots under the P2 gate. The transport measurement then detects the electron current tunneling through the elongated jellybean region under gate P2.

Figure~\ref{fig:device}(c) shows the transport current measurement through the dot channel as a function of the potential on P2. Two distinct charging behaviors are identified, indicating the existence of at least two types of dots under P2: one dot with stronger capacitive coupling (lever arm of 0.052~eV/V) and charging energy of 7.2~meV (grey dashed lines), and another dot with weaker capacitive coupling (lever arm of 0.008 eV/V) and charging energy of 8.9~meV (red dashed lines). 
This two-fold charging behavior is a first indication that the jellybean dot supports separate electronic structures that can be interpreted as artificial molecule.

In order to confirm that the two types of Coulomb diamonds correspond to two different types of dots under P2, we measure charge stability maps using the SET charge sensor. We can determine the location of the dots based on the relative intensities of the SET signal and the capacitive coupling to each of the gates. The map in Fig.~\ref{fig:stability}(a) shows two distinct charge transitions with different capacitive couplings to gate J, suggesting the presence of a dot weakly coupled to J (vertical transitions), and a dot closer to J (tilted transitions). A third dot closer to the RESB gate can be identified from the charge transitions induced by RESB [almost horizontal transitions in Figs.~\ref{fig:stability}(b) and (c)]. We therefore conclude that there are three dots under gate P2: left and right dots which are closer to the side gates J and RESB, respectively, and a dot located in the middle. At sufficiently high $V_{\rm P2}$ in Figs.~\ref{fig:stability}(a) and (b), the three dots merge into one larger dot.

In Fig.~\ref{fig:stability}(d), the charging voltages of the different dots under gate P2 around the biasing point $V_{\rm P2} = 1.9$~V versus various electrostatic gates are plotted (i.e., how large is the voltage change on a particular gate to load an extra electron onto a specific dot). Gate P2 dominates the transitions of all three dots compared to other gates, which confirms that the three dots are all under gate P2. However, the P2 gate is more strongly coupled to the left dot than to the right dot, resulting from the asymmetric biasing and different widths of the J and RESB barriers.
Figure~\ref{fig:stability}(e) depicts a schematic of the capacitive model of the three-dot system and indicates the relative couplings of each of the dots to each of the gates. This system, which is composed of one dot in the middle and one dot on each side, can be interpreted as an artificial molecule.

\begin{figure*}[ht!]
\centering
  \includegraphics[width=\linewidth]{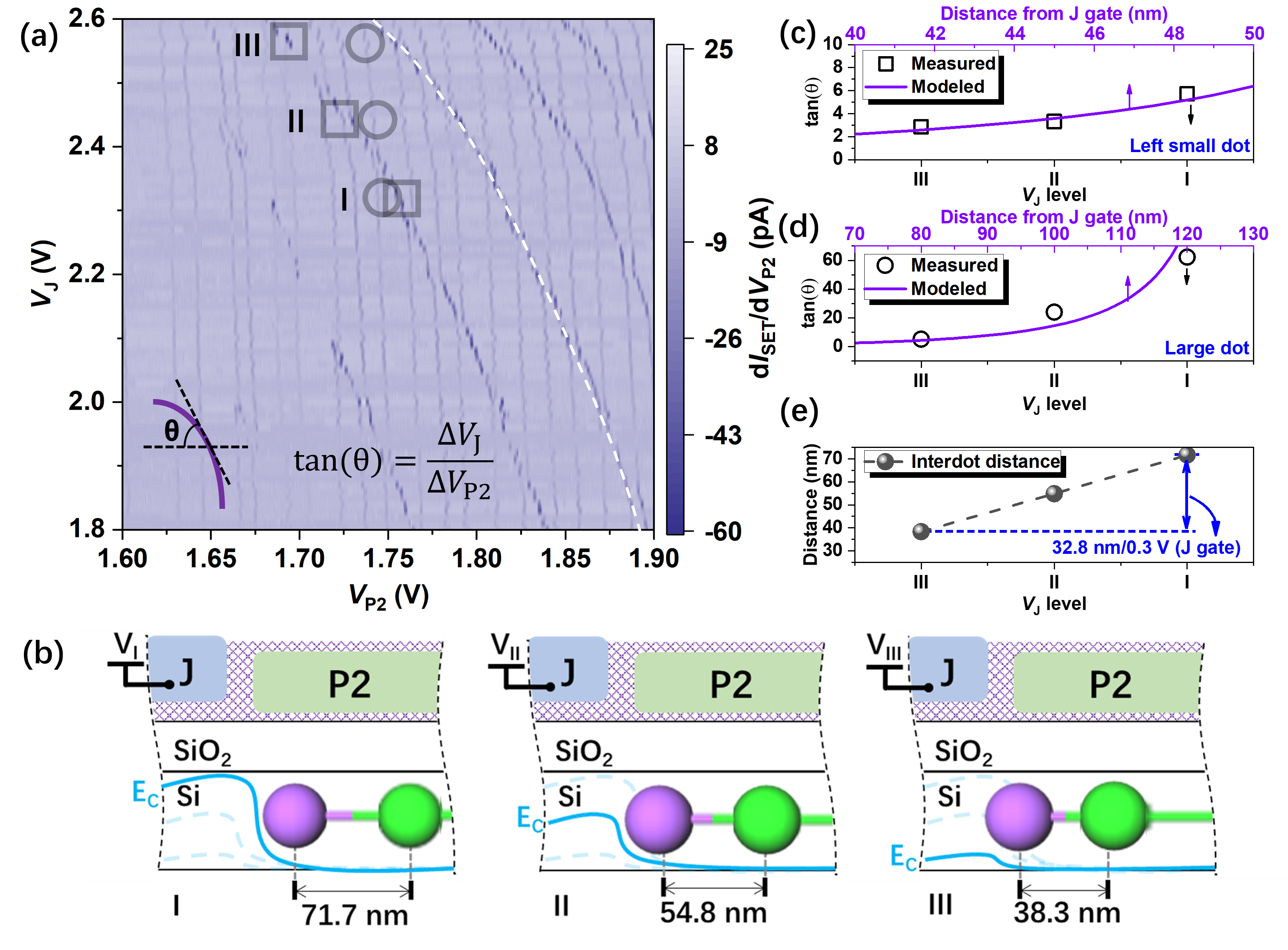}
  \caption{
  Experimentally tuning the distance between artificial atoms. 
  (a) Charge stability map showing the transitions of the dots with bending slopes at high $V_{\rm J}$ (with white dashed line in (a) as a guide to the eye). 
  (b) Schematic showing the shift in dot positions when $V_{\rm J}$ is increased.
  (c)-(e) Relationship between the slope and the position of the dots for (c) the left dot, (d) the middle dot, and (e) the interdot distance.
  }
  \label{fig:tuning}
\end{figure*}

These considerations together with electrostatic simulations (see Sec.~\ref{methods:dotdistance}) allow us to gain an understanding of the experimental system, and how transition lines with different slopes are possible because of the different couplings to each gate. The different couplings are a natural consequence of the P2 dot gate being elongated and 3 times as long as a regular dot gate. The long gate allows for dots to be formed at both edges of the P2 gate that have a stronger coupling to the neighbouring RESB or J gates as compared to the dot that is centered under P2. Furthermore, as shown in the next section, the different gate couplings allow the dots’ relative positions to be shifted by biasing those gates. In other words, the distance between the artificial atoms of the molecule can be tuned.

\section{Tuning of the Artificial Molecule}
We have identified an artificial molecule with three artificial atoms, each coupled to three electrostatic gates with different coupling strengths in the previous section (see Fig.~\ref{fig:stability}). We now show how these differences in coupling strengths can be leveraged to tune the separation between the artificial atoms and, effectively, change the molecular bonds of the artificial molecule. 

\begin{figure*}
\centering
  \includegraphics[width=0.9\linewidth]{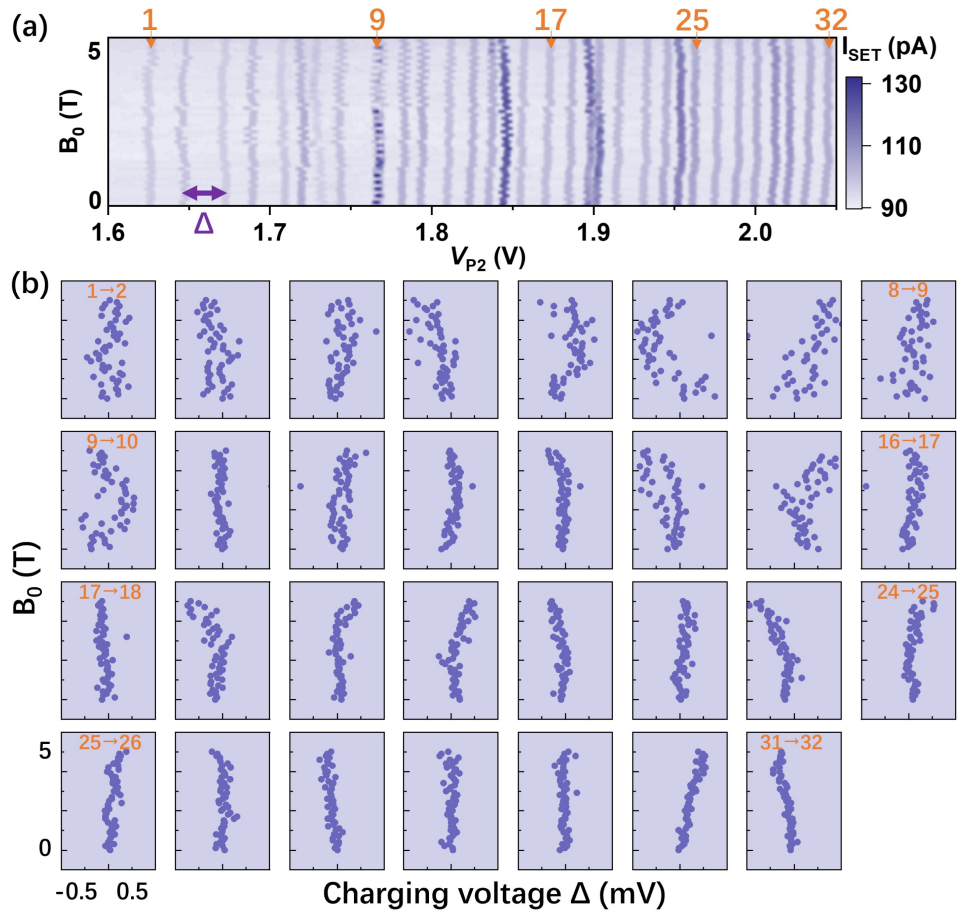}
  \caption{
  Magneto-spectroscopy on the jellybean dot. 
  (a) Transition map of the jellybean dot when sweeping $V_{\rm P2}$ vs. $B_0$. 
  (b) Extracted charging voltage differences $\Delta$ of the first 32 transitions.
  }
  \label{fig:magneto}
\end{figure*}

In Fig.~\ref{fig:tuning}(a), we observe a bending of the transition lines for both the left dot (square markers) and the middle dot (circular markers) for $V_{\rm J}>2.2$~V. This curvature indicates a change in capacitive coupling to the gates, and therefore a shift of the dot positions in a way that is illustrated in Fig.~\ref{fig:tuning}(b). Assuming the charging energy of a dot with a given number of electrons is fixed, the slope of the transition in gate space represents the ratio of the lever arms of the two electrostatic gates and allows us to determine the ratio of the effective coupling capacitances~\cite{mohiyaddin2013noninvasive}. Then, we can model the relationship between the slopes of the lines (note we use ${\rm tan}(\delta)$ here for simplicity) and the dot distance, and plot the extracted distances in Figs.~\ref{fig:tuning}(c) and (d).

With this model, we can determine the positions of the dots from the values of ${\rm tan}(\theta)$. Taking three different voltage levels (I, II, and III) as examples, the left dot shifts from $48.3$~nm (I) to $41.6$~nm (III) distance to the J gate [see Fig.~\ref{fig:tuning}(c)], while the middle dot shifts from $120$~nm (I) to $80$~nm (III) [see Fig.~\ref{fig:tuning}(d)]. Both dots shift closer to the J gate, when its bias is increased. The distance between the two dots is then estimated to change from $\sim70$~nm to $\sim40$~nm, at a rate of $-109$~nm/V [see Fig.~\ref{fig:tuning}(e)]. This change in interdot distance corresponds to a change in the bond length of the artificial molecule, making our system a tunable artificial molecule.

\section{Spin Structure of the Jellybean Dot}

Finally, we conduct a magneto-spectroscopy measurement to show that the jellybean dot has the potential to conserve the spin states while holding or transferring them in future quantum computing architectures. 

In the experiment, we carefully measure the charging voltages as a function of applied magnetic field $B_0$. The voltage differences $\Delta$ between the transitions in Fig.~\ref{fig:magneto}(a) are extracted and plotted in Fig.~\ref{fig:magneto}(b)~\footnote{Unfortunately, the gate stack used in this device (palladium gate electrodes with atomic-layer deposited aluminum oxide) results in much higher charge noise than the previously used aluminum gate stack~\cite{vahapoglu2021coherent}, and renders this measurement noisier than previous measurements.}. Before the 16$^{\rm th}$-to-17$^{\rm th}$ transition, there is no clear spin structure visible, in contrast to what we have previously observed in a circular quantum dot~\cite{leon2020coherent}. This lack of structure is explained by disorder in the device and the complicated interactions between the electrons of the three artificial atoms. In addition, the dot itself is deforming in shape when $V_{\rm J}$ is increased. 

After the 17$^{\rm th}$ transition, the artificial atoms start to merge, leading to a single large dot when the electrostatic potential is dominating over disorder and electron-electron interactions. In this case, the addition of extra electrons follows a more systematic pattern and the magneto-spectroscopy is easier to analyse. While we cannot assign the spin state of the last electron without knowing the spin states of all previous electrons, we can make some interesting observations~\cite{leon2020coherent}. For the 28$^{\rm th}$-29$^{\rm th}$ and 29$^{\rm th}$-30$^{\rm th}$ transitions, the charging voltage $\Delta$ remains constant, which means that the added electron has the same spin state as the previously added electron, indicative of a Hund's rule-like spin filling. For the 18$^{\rm th}$-19$^{\rm th}$, 19$^{\rm th}$-20$^{\rm th}$, and 23$^{\rm rd}$-24$^{\rm th}$ transitions, $\Delta$ shows a clear kink, indicative of an orbital or valley crossing that leads to an electron of a different spin state being loaded for larger $B_0$. 

Overall, the systematic spin filling observed here, promises that a well-defined spin state can be achieved for higher electron occupancies. This should allow for spin shuttling for a spin-1/2 state~\cite{baart2016singlespin,fujita2017coherent,fedele2021simultaneous,sigillito2019siteselective,yoneda2021coherent,noiri2022shuttlingbased}, and jellybean coupling for a spin-0 state~\cite{malinowski2018spin}.

\section{Theoretical Modeling of Spin States in a Jellybean Dot}
\label{sec:theoryhf}

\begin{figure*}[ht!]
\centering
    \includegraphics[width=\linewidth]{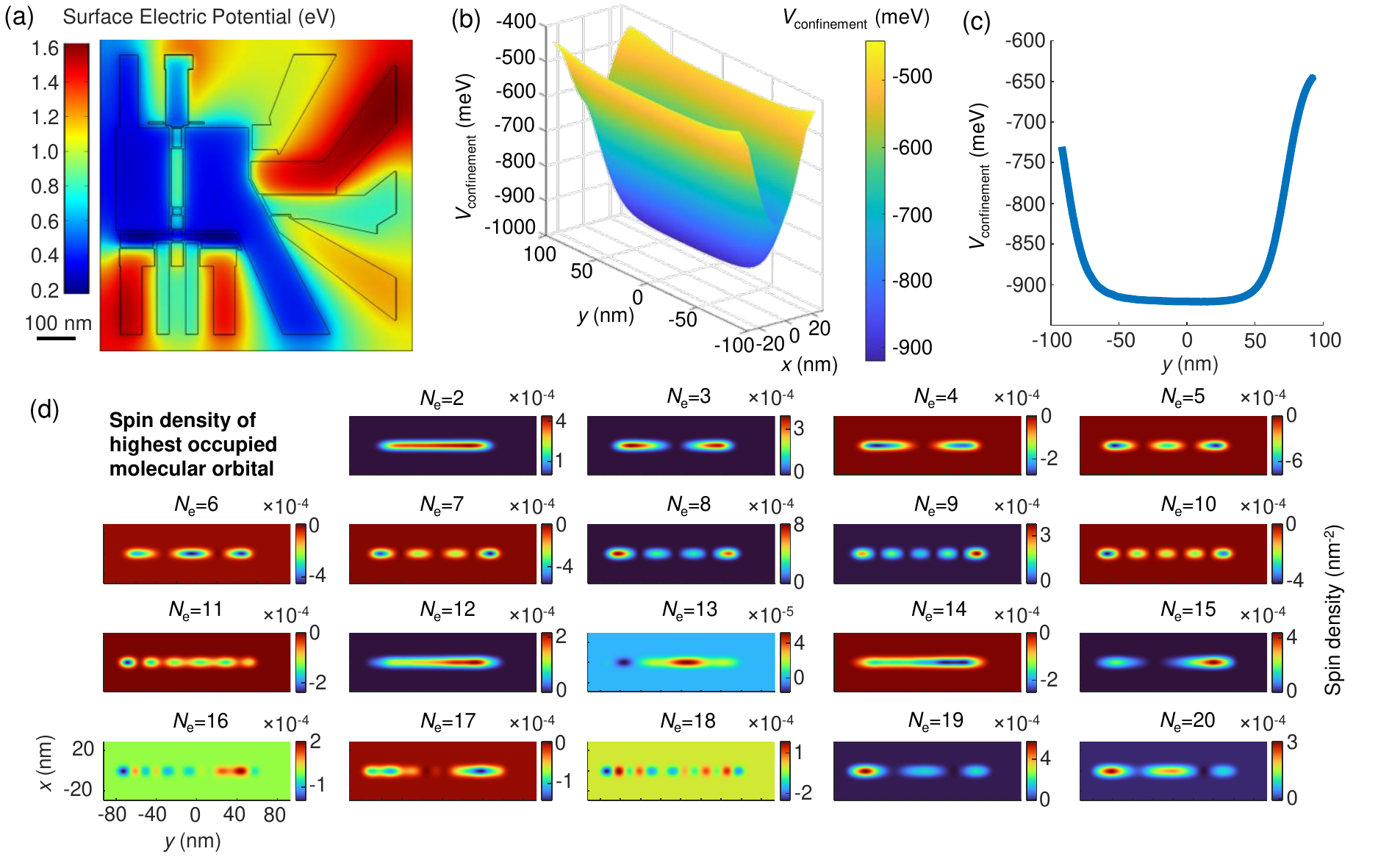}
    \caption{Simulated spin densities from Hartree Fock. 
    (a) 2D cut of the simulated device structure in COMSOL, with the color scale showing the electric potential across the surface of the device. 
    (b) Surface plot of the confinement potential $V_\mathrm{confinement}$ under the P2 gate, which we use to simulate the electrons. 
    (c) 1D cut along the jellybean direction of the confinement potential $V_\mathrm{confinement}$. 
    (d) Spin densities of the highest occupied molecular orbital for systems with varying number of electrons, $N_\mathrm{e}$, from 2 to 20.
    }
  \label{fig:theorymap}
\end{figure*}

We have experimentally shown the charge characteristics of the jellybean dot (Figs.~\ref{fig:device} \& \ref{fig:stability}), and possible signatures of a spin structure (Fig.~\ref{fig:magneto}). In this section, we make use of theoretical modeling to understand the possibility of using such a quantum dot as a mediator for long-range coupling of spin qubits. One of the critical questions with regards to doing so is whether the electronic structure of such an elongated dot is stable and reproducible. In GaAs, the coherent coupling of qubits with mediator dots was successfully demonstrated \cite{croot2018device,malinowski2018spin,malinowski2019fast}, which encourages the investigation of a similar device in MOS silicon. The main scientific questions behind the translation of this technology between platforms are whether 1) the disorder imposed by the SiO$_2$ interface will deter the dot from forming electronic states that extend across the complete dot, and 2) the electron-electron repulsion and quantum correlations in silicon will create significant departures in performance compared to the results in GaAs.

Experimentally, it is hard to distinguish the signatures of 1) and 2). Therefore, we outline below our theoretical model for understanding the different types of electronic states formed at different dot occupancies, highlighting the interaction effects in silicon. We highlight that silicon quantum dots have more potential to form non-trivial quantum states due to electronic interactions than GaAs. This is because silicon has a lower dielectric constant than GaAs, and its conduction band electrons have a larger in-plane effective mass than in GaAs, besides their valley degeneracy (or in general small valley splitting).

Modeling such a multi-electron dot is often difficult because of the large computational load necessary to simulate more than a couple of electrons. In the context of this device, where we have essentially created a multi-dot system under a single large gate (P2), we are able to use the Hartree-Fock method to simulate the system of many electrons in a single potential well. The key advantage of the Hartree-Fock method lies in its ability to simulate high electron numbers in our system within reasonable times, which is not always possible using other methods like full configuration interaction~ \cite{anderson2022highprecision,ercan2021strong}. The trade off here is in the quantitative accuracy of the results, where one of the assumptions central to the method is that the interaction between the electrons is treated as a mean field such that the interaction of each electron with the other electrons depends only on where the electron is located in this field \cite{pratt1956unrestricted}. Regardless, we find that the method is satisfactory for the purpose of achieving a qualitative understanding of the electron system.

We describe here a brief overview of how the method works, and the details will be contained in the Methods section. There are broadly two separate steps in the Hartree-Fock method: first, obtain the approximate potential of the quantum dot under the gates; second, solve for the Hartree-Fock Hamiltonian iteratively such that the energies and wavefunctions obtained from the final solution are consistent with the initial guess. To optimise the electric potential of the system we expect from our devices, we convert the geometry of the device into a 3D model in COMSOL, upon which we can perform electrostatic calculations. This generates a 3D potential profile of the device, which will be used to model the confinement potential under which the electrons sit in the system. This process is elaborated in more detail in the Methods section.

The surface electric potential of the simulated device is shown in Fig.~\ref{fig:theorymap}(a) at a particular gate configuration of  $V_\mathrm{J}=0.6~\mathrm{V}$ and $V_\mathrm{RESB}=1.0~\mathrm{V}$. Of the general potential profile that is generated, we are most concerned with the potential profile within the boundaries of the P2 gate which is the primary gate under which we performed the experiments as indicated in Fig.~\ref{fig:device}. Isolating the relevant part of the potential, we have effectively an elongated single dot quantum well. The profile of this confinement potential, $V_\mathrm{confinement}$, is shown in Fig.~\ref{fig:theorymap}(b). It can be observed that the potential is approximately parabolic along the $x$ direction but is more like a box potential along the $y$ direction. To highlight the profile of the confinement potential in the $y$ direction, we plot a cut along $y$ at $x=0~\mathrm{nm}$ in Fig.~\ref{fig:theorymap}(c). The obtained potential is then used to perform the Hartree-Fock algorithm.

The Hartree-Fock method begins by guessing an initial value for the energy and wavefunctions, which can be calculated based on the single electron Hamiltonian without interaction or we can use a previous solution. We define the system with a specific number of electrons and construct the Hamiltonian for the system by discretizing the 3D simulation cell defining the quantum dot, similar to a tight-binding model. We are essentially performing general unrestricted Hartree-Fock theory where we are minimising the energy of a single Slater determinant in the most energetically favourable configuration, which does not restrict the spin degree of freedom. With each iteration of the method, there is a small rearrangement of the charge density of the electrons in the system as a function of position and the total energy also changes. Eventually, we consider the solution to be converged when the change in total energy, charge density, and exchange energy are respectively smaller than $10^{-4}$~meV, $10^{-4}$~meV, and $10^{-3}$~meV.

After obtaining a converged solution to the Hartree-Fock Hamiltonian, the two main outputs of interest are the total energies of the multi-electron system and the wavefunctions of the electrons in the system. We first examine the wavefunction outputs by looking at specifically the results obtained from the confinement potential defined by $V_\mathrm{J}=0.6~\mathrm{V}$, $V_\mathrm{RESB}=1.0~\mathrm{V}$, and $V_\mathrm{P2}=1.8~\mathrm{V}$. Using the output wavefunctions, we are able to calculate the spin densities of the highest occupied molecular orbital,
\begin{align*}
    \rho_{i,\mathrm{spin}}(x,y) &= \sum_{z} \rho_{i,\uparrow}(x,y,z) - \rho_{i,\downarrow}(x,y,z) \\
    &= |\psi_{\uparrow}(x,y,z)|^2 - |\psi_{\downarrow}(x,y,z)|^2,
    \label{eq:spindensity}
\end{align*}
where these densities are summed along the $z$ direction. The spin density informs us of the spin orbital shapes and the spin state of the electron in the highest occupied molecular orbitals, where a positive and negative density indicate spin up and spin down states, respectively. We note that because we are using an unrestricted Hartree-Fock algorithm, which does not place constraints on the spin states of the solution, the spin up and down wavefunctions are minimised separately \cite{pratt1956unrestricted}.

In Fig.~\ref{fig:theorymap}(d), we plot the spin densities, extracted for the jellybean quantum dot being occupied with 2 to 20 electrons.  The results show that some of the occupation numbers form non-trivial states where charge localises in a chain, while other occupations reveal a smoother charge density distribution spanning the full quantum dot. Our results corroborate theoretically our earlier hypothesis about the physics of such a jellybean dot in silicon, and the non-negligible, strong electron-electron interactions. In these simulations, we consider primarily the effect of electron-electron interaction to attain an understanding of the system and to show that electron-electron interaction alone would be capable of causing this formation of non-trivial states. In further studies, we could also account for effects of surface roughness and explicit spin-orbit coupling terms.%~\cite{cifuentes2022roughness}.

\section{Conclusion}
In this article, we demonstrated charge transport through and charging of an elongated quantum dot -- a so-called jellybean quantum dot -- where the gate is about five times longer than for a typical quantum dot. The data shows that disorder and electron-electron interaction lead to the formation of one dot under the middle of the gate and two dots at the ends of the gate, corresponding to the formation of a tunable, artificial molecule. 
This interpretation is corroborated with electrostatic simulations in COMSOL and Hartree-Fock multi-electron simulations for up to twenty electrons. 
Finally, we presented magneto-spectroscopy results, which are indicative of well-defined spin states of the jellybean quantum dot formed in the device.

Our results are a first demonstration of the jellybean structure in SiMOS architectures, and will be instrumental in understanding the physics of jellybean couplers in silicon as we move towards using the jellybean as a inter-qubit mediator of exchange coupling.

%%%%%%%%%%%%%%%%%%%%%%%%%%%%%%%%%%%%%%%%%%%%%%%%%%%%%%%%%
\section{Methods}

\subsection{Measurement Setup}
The device was measured in a Bluefors LD400 dilution refrigerator with a base temperature of $T_{\rm MXC}\approx 22$~mK. The DC magnetic field was applied with an American Magnetics 5~T superconducting split-coil magnet. DC bias voltages were generated from QDAC high-precision low-noise computer-controlled voltage generator. The SET current and the charge transition of device were measured by a double lock-in technique \cite{Yang2011dynamically}.

\subsection{Hartree-Fock Method}
\label{methods:hartreefock}
In Fig.~\ref{fig:theorymap}, we showed results obtained from electrostatic simulations performed using the Hartree-Fock method. In order to perform these simulations, there are several input parameters that will need to be included. We begin with the confinement potential, $V_\mathrm{confinement}$ which defines the potential in which the electrons sit. Due to the size of the dot gate for the jellybean dot, we expect that the potential may not be defined by a typical harmonic potential well. Therefore, we use COMSOL to simulate the potential of the jellybean dot device. We begin by constructing a realistic 3D model of the device as shown in Fig.~\ref{fig:device}(a) using COMSOL, with one of the outputs being a realistic potential that describes accurately the structure of the device, including but not limited to the substrate and the expected formation of the different gate layers during the fabrication process.

Taking this model, we are able to generate in COMSOL the potential profile that is formed under the gates given a particular set of gate voltages using the electrostatics module. We are able to emulate the voltage configuration used in the experiment and approximately define the shape of the potential, and also generate potentials while simulating accurately the effects of the gates on the device. We use these simulations to obtain a set of confinement potentials that are defined by the P2 gate, the J gate, and the RESB gate, all of which are labelled in Fig.~\ref{fig:device}(a).

We will now detail the Hartree-Fock method as it is performed in our simulations, which more commonly would be considered as the unrestricted Hartree-Fock method \cite{pratt1956unrestricted}. At its core, the simulation works by considering all electrons other than the electron being studied as a static electric field, and considers the effect of that field on the electron. The system in which the electrons interact is defined by two properties in our simulations. The first is the confinement potential, which we have just described. The second is the simulation cell in which we perform the simulations. The size of the simulation cell is strongly correlated with the potential profile obtained from the COMSOL simulations, and is defined such that we include only the area defined by the P2 dot gate, and does not extend into neighboring gates. 

We also consider in our calculations the charge, spin, and valley degrees of freedom. The charge degree of freedom is defined by two quantities; one is the number of electrons in our system, and the other is the location of the electrons within the simulation cell. The spin degree of freedom is defined such that we have two separate wavefunctions describing the spin up and down states. Finally, the valley degree is defined by modeling different hopping parameters along the interface direction \cite{ercan2021strong}. In the following paragraphs, we describe the mathematical model for the Hartree-Fock method.

The complete Hartree-Fock Hamiltonian is given by,
\begin{align}
    H_\mathrm{HF} = \sum_i H_\mathrm{single}(\mathbf{r}_i) + H_\mathrm{int}(\mathbf{r}_1,...,\mathrm{r}_N),
\end{align}
where $H_\mathrm{single}(\mathbf{r}_i)$ is the single electron Hamiltonian, without considering electron-electron interaction, and can be written as follows,
\begin{align}
    H_\mathrm{single} = H_\mathrm{K} + V_\mathrm{c},
\end{align}
where $H_\mathrm{K}$ can be interpreted as the kinetic energy of the system and $V_\mathrm{c}$ is the confinement potential of the system, including both the quantum well potential and the vertical confinement electric field.

We construct the Hamiltonian by considering a cuboid simulation cell in three dimensions, with a fourth dimension accounting for spin, effectively representing the Hamiltonian using a 4D matrix. We divide the cuboid into grids along each of the Cartesian directions, and we choose grid sizes of 0.4 nm, 0.8 nm and 0.1 nm along the $x$, $y$, and $z$ directions. The grid sizes are chosen such that they are small enough to account accurately for changes in the wavefunction in space, while also not too small that it becomes computationally difficult. This is because the size of the Hamiltonian is defined by the size of the cuboid grid along the Cartesian directions, and the last dimension of spin having a size of 2.

In essence, the Hartree-Fock algorithm takes into account electron-electron interactions in addition to the Fock states (comprising the kinetic and potential terms) in constructing the Hamiltonian. We use a self-consistent field method to solve the system Hamiltonian iteratively to minimise the total energy of the system. This method allows us to simulate large numbers of electrons in our system, as demonstrated in the main result of the paper. 

We design our method following a similar vein to that in Ref.~\onlinecite{ercan2021strong}, which considers a tight-binding model of electrons and accounts for the valley degree of freedom by having different hopping coefficients for different sites along the $z$ direction. Accounting for the valley degree of freedom is a key ingredient that allows us to account for the distribution of electrons accurately in silicon-based devices.

One of the main factors affecting the accuracy and efficiency of the method is the starting point of the algorithm. In order to estimate a starting point as close to the final solution as possible without having to actually perform the algorithm, we calculate the single particle states of the system. We consider an analytical potential model that is close to the expected potential extracted from COMSOL simulations and use that to calculate the single particle states. The single particle states will be used as the initial point of the simulation. 

As a starting point for the simulations, we solve for the single particle Hamiltonian,
\begin{align}
    H_\mathrm{single} = H_\mathrm{K} + V_\mathrm{pot},
\end{align}
where $H_\mathrm{K}$ is the kinetic energy Hamiltonian and $V_\mathrm{pot}$ refers to the confinement potential as described. Without taking into account the electron-electron repulsion, we are able to calculate the single particle state energies and that serves as the starting point of the simulation. This ensures that the initial state is sufficiently close to the final state.

During the method, the total energies of the electrons in the system will be minimised and is given by,
\begin{align}
    \mathcal{E} = \sum_i \varepsilon_i - \frac{1}{2}\sum_j V_\mathrm{H} \sum_i |\psi_i|^2 - V_\mathrm{Ex},
\end{align}
where sums over $i$ are over the total number of electrons while sums with respect to $j$ are over the total space. We note here that these are essentially the Fock terms, the Hartree terms and the exchange term, respectively. We then calculate the change in the total energy by comparing it with the result obtained in the previous iteration. We consider the energy as converged when the change in total energy, $\Delta E$, is less than $10^{-4}~\mathrm{meV}$. This is the first check for convergence as part of a three-step check before the algorithm deems the solution to be converged. Should the total energy, $E$, be converged, the algorithm will then check for the change in wavefunction at this iteration step, $\Delta\psi$, and will be deemed as converged if it is less than $10^{-4}~\mathrm{meV}$. Finally, if the the wavefunction is converged, the algorithm will re-calculate the exchange energy for the new electronic configuration and check for the change in the exchange energy, $\Delta E_\mathrm{ex}$, and the solution will be considered overall converged, if this change is less than $10^{-3}~\mathrm{meV}$. This is based on the idea that the initial calculation of exchange may be wrong and that would also result in an erroneous result for the wavefunction and energy, and if that is not the case, the wavefunction and the exchange energy would be consistent and therefore leading to a small $\Delta E_\mathrm{ex}$. By defining convergence in this way, we can afford to set looser bounds on the convergence criteria, and even though there are three different sets of convergence criteria that needs to be met, there exists a speed-up in the algorithm due to the looser bounds. If the energy and wavefunction convergences are not achieved, the wavefunction will be mixed to include a proportion of the newly calculated wavefunction in the following way,
\begin{align}
    \psi_\mathrm{new} = (1-\beta)\psi_{i-1} + \beta*\psi_{i},
\end{align}
where $\beta$ typically varies from $10^{-3}$ to $10^{-1}$. Should $\Delta E$ and $\Delta\psi$ both be $<10^{-4}~\mathrm{meV}$, but the re-calculated exchange energy differs from its initial value by a significant amount larger than $10^{-3}$~meV, the algorithm re-approaches the converged solution with this new value for exchange energy. This method of converging the exchange energy first reduces the overall time required to perform the algorithm as the calculation for the exchange operator can take up a significant portion of the total run time. Second, this stabilises the algorithm as it ensures that not too many parameters in the simulation are being changed at each stage of the simulation.

After convergence is achieved, we store the two primary outputs of the Hartree-Fock method, namely the total energy of the system, and the wavefunctions of the electrons in the system. The spin densities can be calculated by taking the difference between the highest occupied molecular orbital spin up and down densities as shown in Eq.~\ref{eq:spindensity}.

We will also emphasise that our simulations are only accurate up to the accuracy of the set convergence criteria. There remains error on the order of magnitude smaller than $10^{-4}~\mathrm{meV}$ for the energy and wavefunction, as well as errors smaller than $10^{-3}~\mathrm{meV}$ for the exchange energy. Therefore, the accuracy of the wavefunctions can vary and some inconsistencies as well as artefacts in the wavefunctions can remain at this level of accuracy. We are careful to examine the wavefunction only on its general shape and not its detailed structure at every grid position.

\begin{figure*}
\centering
  \includegraphics[width=0.75\linewidth]{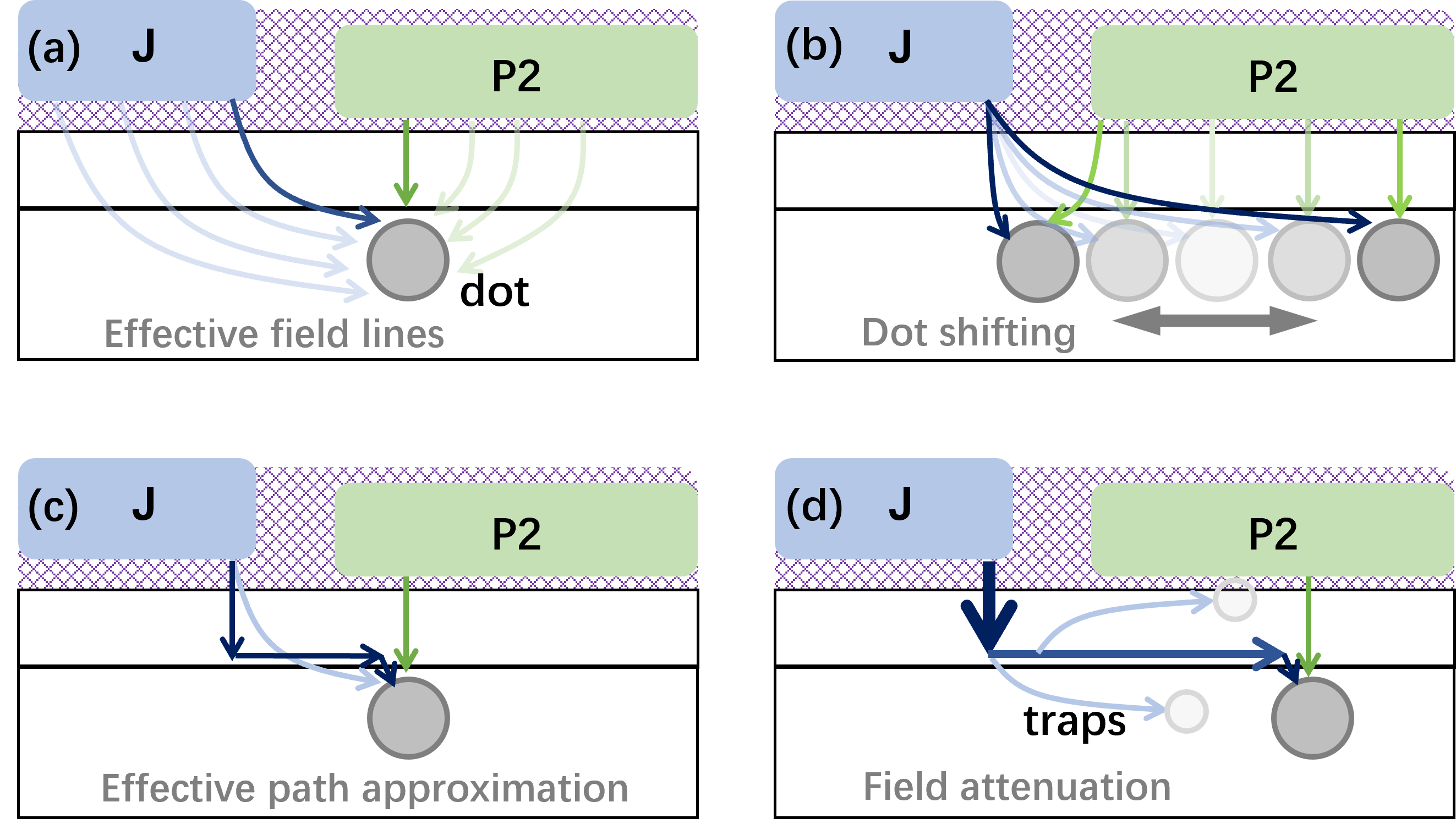}
  \caption{
  The modeling of the electrically tunable artificial molecule. 
  (a) The electric field distribution is simplified as effective field lines, where the dark colored arrows are the simplified electric field lines.  
  (b) Illustration of shifting the dot position. 
  (c) Simplification of the trace of the electric field lines.
  (d) Charge traps will disperse the strength of the electric field line.
  }
  \label{fig:dotdistmodel}
\end{figure*}

\subsection{Dot distance modelling}
\label{methods:dotdistance}

For the dot distance calculations, we employ a simplified field-effect model, which is depicted in Fig.~\ref{fig:dotdistmodel}.
We assume that there is only one main electric field line that represents the real influence of the complicated electric field distributions, shown as solid dark color arrows in Figs.~\ref{fig:dotdistmodel}(a) and (b). 
If the displacement distance of the dot is small enough, we can then simplify the electric field line’s trace into two straight lines as Fig.~\ref{fig:dotdistmodel}(c), which corresponds to the so-called effective path approximation. This simplification is valid if the P2 voltage is similar to the J gate voltage as in Fig.~\ref{fig:tuning}. 
However, owing to the traps that would disperse the electric field, the effective electric field strength will decrease along the direction it extends. We, therefore, need to include some compensation to reduce the error further. The detailed expression of the model can be found in the following paragraphs.

To simplify the calculation, we treat the dot under P2 as a point charge. The capacitive coupling between a specific gate and the dot can then be depicted by

\begin{align}
C_{\rm tot} = \sum_\mathcal{L} \frac{p_\mathcal{L}}{\int_\mathbb{C}\frac{d\vec{r}}{\epsilon_\mathcal{L}(\vec{r})}},
\end{align}

where $\vec{r}$ is the unit vector along the electric field line $\mathbb{C}$, and $p_\mathcal{L}$ and $\epsilon_\mathcal{L}$ are the weight of the coupling strength and the local permittivity of the path $\mathcal{L}$, respectively. When the P2 voltage is high, the electric field lines from the J gate are likely to be concave, and those from the P2 gate are likely to be straight lines because of the relative location between the dot and the P2 gate, as schematically shown in Figs.~\ref{fig:dotdistmodel}(a) and (b). In this case, the main part of the electric field lines emitted from a specific gate can be condensed into one representative path, e.g., $\mathcal{L}^{'}$ and $\mathcal{L}^{''}$ for the J-dot and P2-dot path, respectively. Then, the capacitance between the gate and the dot can be expressed as 

\begin{align}
C_{\rm J-dot} = \int_\mathbb{C} \frac{\epsilon_{\mathcal{L}'}(\vec{r})} {d\vec{r}},
\end{align}

and

\begin{align}
C_{\rm P2-dot} = \int_\mathbb{C} \frac{\epsilon_{\mathcal{L}''}(\vec{r})} {d\vec{r}}.
\end{align}

When sweeping the J (P2) gate by the amount of the charging voltage $\Delta V_{\rm J}$ ($\Delta V_{\rm P2}$), as is done in the measurement in Fig.~\ref{fig:tuning}(a), the charge of the dot changes by one electron. So, the ratio of charging voltages $\tan\theta$ also equals

\begin{align}
\frac{\Delta{V_{\rm J}}}{\Delta{V_{\rm P2}}} = \frac{\int_{\mathbb{C}_{\rm P2}} \frac{\epsilon_{\mathcal{L}''}(\vec{r})} {d\vec{r}}} {\int_{\mathbb{C}_{\rm J}} \frac{\epsilon_{\mathcal{L}'}(\vec{r})} {d\vec{r}}}.
\end{align}

Here, the influence from the anisotropy of the dielectrics can be reasonably neglected for simplification. Additionally, by assuming the paths are polylines instead of curves, as shown in Fig.~\ref{fig:dotdistmodel}(c), the ratio can be further simplified as

\begin{align}
\tan{\theta} = \frac{\sum_i\frac{\epsilon_i}{t_i^{\rm eff}}}{\sum_i\frac{\epsilon_j}{t_j^{\rm eff}}},
\end{align}

where i and j represent the different dielectrics along the path and $t^{\rm eff}$ is the effective thickness of the dielectric along the path. For the P2 gate, the dot is near so that the attenuation imposed by the charge traps can be neglected. However, for the J gate, which is far from the dot, the attenuation is no longer negligible, as shown in Fig.~\ref{fig:dotdistmodel}(d). In this work, it is assumed to be a factor that increases the intrinsic thickness of the dielectric $t^{\rm int}$ and is calculated in the following form~\cite{kindt1998modelling,botkin2004mathematical}:

\begin{align}
t^{\rm eff} = t^{\rm int} \cdot e^{\frac{\left|\vec{\ell}\right|-\ell_0}{\ell_0}},
\end{align}

where $\vec{\ell}$ represents the current position of the dot and $\ell_0$ is the initial distance between the dot and the J gate. Referring to the fabrication layout design, we estimate the left dot and the middle dot are 45~nm and 130~nm away from the middle of the J gate, respectively. Then, given the thickness of each ALD-grown dielectric layer, the relationship between the dot’s location and the charging voltage ratio can be achieved, as articulated in the main text.

% Acknowledgements
\medskip
\textbf{Acknowledgements} \par %delete if not applicable))
We acknowledge support from the Australian Research Council (FL190100167, CE170100012 and LE160100069), the US Army Research Office (W911NF-17-1-0198), the NSW Node of the Australian National Fabrication Facility and UNSW Sydney. The views and conclusions contained in this document are those of the authors and should not be interpreted as representing the official policies, either expressed or implied, of the Army Research Office or the US Government. The US Government is authorized to reproduce and distribute reprints for Government purposes notwithstanding any copyright notation herein. Z.W., M.K.F., S.S., W.G., J.Y.H., Y.S. acknowledge support from the Sydney Quantum Academy. A.L. acknowledges support through the UNSW Scientia program.

% References
\medskip
\bibliography{main}

%Control: production of eprint (1) enabled
\begin{thebibliography}{50}%
\makeatletter
\providecommand \@ifxundefined [1]{%
 \@ifx{#1\undefined}
}%
\providecommand \@ifnum [1]{%
 \ifnum #1\expandafter \@firstoftwo
 \else \expandafter \@secondoftwo
 \fi
}%
\providecommand \@ifx [1]{%
 \ifx #1\expandafter \@firstoftwo
 \else \expandafter \@secondoftwo
 \fi
}%
\providecommand \natexlab [1]{#1}%
\providecommand \enquote  [1]{``#1''}%
\providecommand \bibnamefont  [1]{#1}%
\providecommand \bibfnamefont [1]{#1}%
\providecommand \citenamefont [1]{#1}%
\providecommand \href@noop [0]{\@secondoftwo}%
\providecommand \href [0]{\begingroup \@sanitize@url \@href}%
\providecommand \@href[1]{\@@startlink{#1}\@@href}%
\providecommand \@@href[1]{\endgroup#1\@@endlink}%
\providecommand \@sanitize@url [0]{\catcode `\\12\catcode `\$12\catcode
  `\&12\catcode `\#12\catcode `\^12\catcode `\_12\catcode `\%12\relax}%
\providecommand \@@startlink[1]{}%
\providecommand \@@endlink[0]{}%
\providecommand \url  [0]{\begingroup\@sanitize@url \@url }%
\providecommand \@url [1]{\endgroup\@href {#1}{\urlprefix }}%
\providecommand \urlprefix  [0]{URL }%
\providecommand \Eprint [0]{\href }%
\providecommand \doibase [0]{https://doi.org/}%
\providecommand \selectlanguage [0]{\@gobble}%
\providecommand \bibinfo  [0]{\@secondoftwo}%
\providecommand \bibfield  [0]{\@secondoftwo}%
\providecommand \translation [1]{[#1]}%
\providecommand \BibitemOpen [0]{}%
\providecommand \bibitemStop [0]{}%
\providecommand \bibitemNoStop [0]{.\EOS\space}%
\providecommand \EOS [0]{\spacefactor3000\relax}%
\providecommand \BibitemShut  [1]{\csname bibitem#1\endcsname}%
\let\auto@bib@innerbib\@empty
%</preamble>
\bibitem [{\citenamefont {DiVincenzo}(2000)}]{divincenzo2000physical}%
  \BibitemOpen
  \bibfield  {author} {\bibinfo {author} {\bibfnamefont {D.~P.}\ \bibnamefont
  {DiVincenzo}},\ }\bibfield  {title} {\bibinfo {title} {The physical
  implementation of quantum computation},\ }\href
  {https://doi.org/10.1002/1521-3978(200009)48:9/11<771::AID-PROP771>3.0.CO;2-E}
  {\bibfield  {journal} {\bibinfo  {journal} {Fortschritte der Physik}\
  }\textbf {\bibinfo {volume} {48}},\ \bibinfo {pages} {771} (\bibinfo {year}
  {2000})}\BibitemShut {NoStop}%
\bibitem [{\citenamefont {Zhang}\ \emph {et~al.}(2018)\citenamefont {Zhang},
  \citenamefont {Li}, \citenamefont {Wang}, \citenamefont {Cao}, \citenamefont
  {Xiao},\ and\ \citenamefont {Guo}}]{zhang2018qubits}%
  \BibitemOpen
  \bibfield  {author} {\bibinfo {author} {\bibfnamefont {X.}~\bibnamefont
  {Zhang}}, \bibinfo {author} {\bibfnamefont {H.-O.}\ \bibnamefont {Li}},
  \bibinfo {author} {\bibfnamefont {K.}~\bibnamefont {Wang}}, \bibinfo {author}
  {\bibfnamefont {G.}~\bibnamefont {Cao}}, \bibinfo {author} {\bibfnamefont
  {M.}~\bibnamefont {Xiao}},\ and\ \bibinfo {author} {\bibfnamefont {G.-P.}\
  \bibnamefont {Guo}},\ }\bibfield  {title} {\bibinfo {title} {Qubits based on
  semiconductor quantum dots},\ }\href
  {https://doi.org/10.1088/1674-1056/27/2/020305} {\bibfield  {journal}
  {\bibinfo  {journal} {Chinese Physics B}\ }\textbf {\bibinfo {volume} {27}},\
  \bibinfo {pages} {020305} (\bibinfo {year} {2018})}\BibitemShut {NoStop}%
\bibitem [{\citenamefont {Laucht}\ \emph {et~al.}(2021)\citenamefont {Laucht},
  \citenamefont {Hohls}, \citenamefont {Ubbelohde}, \citenamefont
  {{Gonzalez-Zalba}}, \citenamefont {Reilly}, \citenamefont {Stobbe},
  \citenamefont {Schr{\"o}der}, \citenamefont {Scarlino}, \citenamefont
  {Koski}, \citenamefont {Dzurak}, \citenamefont {Yang}, \citenamefont
  {Yoneda}, \citenamefont {Kuemmeth}, \citenamefont {Bluhm}, \citenamefont
  {Pla}, \citenamefont {Hill}, \citenamefont {Salfi}, \citenamefont {Oiwa},
  \citenamefont {Muhonen}, \citenamefont {Verhagen}, \citenamefont {LaHaye},
  \citenamefont {Kim}, \citenamefont {Tsen}, \citenamefont {Culcer},
  \citenamefont {Geresdi}, \citenamefont {Mol}, \citenamefont {Mohan},
  \citenamefont {Jain},\ and\ \citenamefont {Baugh}}]{laucht2021roadmap}%
  \BibitemOpen
  \bibfield  {author} {\bibinfo {author} {\bibfnamefont {A.}~\bibnamefont
  {Laucht}}, \bibinfo {author} {\bibfnamefont {F.}~\bibnamefont {Hohls}},
  \bibinfo {author} {\bibfnamefont {N.}~\bibnamefont {Ubbelohde}}, \bibinfo
  {author} {\bibfnamefont {M.~F.}\ \bibnamefont {{Gonzalez-Zalba}}}, \bibinfo
  {author} {\bibfnamefont {D.~J.}\ \bibnamefont {Reilly}}, \bibinfo {author}
  {\bibfnamefont {S.}~\bibnamefont {Stobbe}}, \bibinfo {author} {\bibfnamefont
  {T.}~\bibnamefont {Schr{\"o}der}}, \bibinfo {author} {\bibfnamefont
  {P.}~\bibnamefont {Scarlino}}, \bibinfo {author} {\bibfnamefont {J.~V.}\
  \bibnamefont {Koski}}, \bibinfo {author} {\bibfnamefont {A.}~\bibnamefont
  {Dzurak}}, \bibinfo {author} {\bibfnamefont {C.-H.}\ \bibnamefont {Yang}},
  \bibinfo {author} {\bibfnamefont {J.}~\bibnamefont {Yoneda}}, \bibinfo
  {author} {\bibfnamefont {F.}~\bibnamefont {Kuemmeth}}, \bibinfo {author}
  {\bibfnamefont {H.}~\bibnamefont {Bluhm}}, \bibinfo {author} {\bibfnamefont
  {J.}~\bibnamefont {Pla}}, \bibinfo {author} {\bibfnamefont {C.}~\bibnamefont
  {Hill}}, \bibinfo {author} {\bibfnamefont {J.}~\bibnamefont {Salfi}},
  \bibinfo {author} {\bibfnamefont {A.}~\bibnamefont {Oiwa}}, \bibinfo {author}
  {\bibfnamefont {J.~T.}\ \bibnamefont {Muhonen}}, \bibinfo {author}
  {\bibfnamefont {E.}~\bibnamefont {Verhagen}}, \bibinfo {author}
  {\bibfnamefont {M.~D.}\ \bibnamefont {LaHaye}}, \bibinfo {author}
  {\bibfnamefont {H.~H.}\ \bibnamefont {Kim}}, \bibinfo {author} {\bibfnamefont
  {A.~W.}\ \bibnamefont {Tsen}}, \bibinfo {author} {\bibfnamefont
  {D.}~\bibnamefont {Culcer}}, \bibinfo {author} {\bibfnamefont
  {A.}~\bibnamefont {Geresdi}}, \bibinfo {author} {\bibfnamefont {J.~A.}\
  \bibnamefont {Mol}}, \bibinfo {author} {\bibfnamefont {V.}~\bibnamefont
  {Mohan}}, \bibinfo {author} {\bibfnamefont {P.~K.}\ \bibnamefont {Jain}},\
  and\ \bibinfo {author} {\bibfnamefont {J.}~\bibnamefont {Baugh}},\ }\bibfield
   {title} {\bibinfo {title} {Roadmap on quantum nanotechnologies},\ }\href
  {https://doi.org/10.1088/1361-6528/abb333} {\bibfield  {journal} {\bibinfo
  {journal} {Nanotechnology}\ }\textbf {\bibinfo {volume} {32}},\ \bibinfo
  {pages} {162003} (\bibinfo {year} {2021})}\BibitemShut {NoStop}%
\bibitem [{\citenamefont {Chatterjee}\ \emph {et~al.}(2021)\citenamefont
  {Chatterjee}, \citenamefont {Stevenson}, \citenamefont {De~Franceschi},
  \citenamefont {Morello}, \citenamefont {{de Leon}},\ and\ \citenamefont
  {Kuemmeth}}]{chatterjee2021semiconductor}%
  \BibitemOpen
  \bibfield  {author} {\bibinfo {author} {\bibfnamefont {A.}~\bibnamefont
  {Chatterjee}}, \bibinfo {author} {\bibfnamefont {P.}~\bibnamefont
  {Stevenson}}, \bibinfo {author} {\bibfnamefont {S.}~\bibnamefont
  {De~Franceschi}}, \bibinfo {author} {\bibfnamefont {A.}~\bibnamefont
  {Morello}}, \bibinfo {author} {\bibfnamefont {N.~P.}\ \bibnamefont {{de
  Leon}}},\ and\ \bibinfo {author} {\bibfnamefont {F.}~\bibnamefont
  {Kuemmeth}},\ }\bibfield  {title} {\bibinfo {title} {Semiconductor qubits in
  practice},\ }\href {https://doi.org/10.1038/s42254-021-00283-9} {\bibfield
  {journal} {\bibinfo  {journal} {Nature Reviews Physics}\ }\textbf {\bibinfo
  {volume} {3}},\ \bibinfo {pages} {157} (\bibinfo {year} {2021})}\BibitemShut
  {NoStop}%
\bibitem [{\citenamefont {{Gonzalez-Zalba}}\ \emph {et~al.}(2021)\citenamefont
  {{Gonzalez-Zalba}}, \citenamefont {{de Franceschi}}, \citenamefont {Charbon},
  \citenamefont {Meunier}, \citenamefont {Vinet},\ and\ \citenamefont
  {Dzurak}}]{gonzalez-zalba2021scaling}%
  \BibitemOpen
  \bibfield  {author} {\bibinfo {author} {\bibfnamefont {M.~F.}\ \bibnamefont
  {{Gonzalez-Zalba}}}, \bibinfo {author} {\bibfnamefont {S.}~\bibnamefont {{de
  Franceschi}}}, \bibinfo {author} {\bibfnamefont {E.}~\bibnamefont {Charbon}},
  \bibinfo {author} {\bibfnamefont {T.}~\bibnamefont {Meunier}}, \bibinfo
  {author} {\bibfnamefont {M.}~\bibnamefont {Vinet}},\ and\ \bibinfo {author}
  {\bibfnamefont {A.~S.}\ \bibnamefont {Dzurak}},\ }\bibfield  {title}
  {\bibinfo {title} {Scaling silicon-based quantum computing using {CMOS}
  technology},\ }\href {https://doi.org/10.1038/s41928-021-00681-y} {\bibfield
  {journal} {\bibinfo  {journal} {Nature Electronics}\ }\textbf {\bibinfo
  {volume} {4}},\ \bibinfo {pages} {872} (\bibinfo {year} {2021})}\BibitemShut
  {NoStop}%
\bibitem [{\citenamefont {Itoh}\ and\ \citenamefont
  {Watanabe}(2014)}]{itoh2014isotope}%
  \BibitemOpen
  \bibfield  {author} {\bibinfo {author} {\bibfnamefont {K.~M.}\ \bibnamefont
  {Itoh}}\ and\ \bibinfo {author} {\bibfnamefont {H.}~\bibnamefont
  {Watanabe}},\ }\bibfield  {title} {\bibinfo {title} {Isotope engineering of
  silicon and diamond for quantum computing and sensing applications},\ }\href
  {https://doi.org/10.1557/mrc.2014.32} {\bibfield  {journal} {\bibinfo
  {journal} {MRS Communications}\ }\textbf {\bibinfo {volume} {4}},\ \bibinfo
  {pages} {143} (\bibinfo {year} {2014})}\BibitemShut {NoStop}%
\bibitem [{\citenamefont {Saraiva}\ \emph {et~al.}(2022)\citenamefont
  {Saraiva}, \citenamefont {Lim}, \citenamefont {Yang}, \citenamefont {Escott},
  \citenamefont {Laucht},\ and\ \citenamefont {Dzurak}}]{saraiva2022materials}%
  \BibitemOpen
  \bibfield  {author} {\bibinfo {author} {\bibfnamefont {A.}~\bibnamefont
  {Saraiva}}, \bibinfo {author} {\bibfnamefont {W.~H.}\ \bibnamefont {Lim}},
  \bibinfo {author} {\bibfnamefont {C.~H.}\ \bibnamefont {Yang}}, \bibinfo
  {author} {\bibfnamefont {C.~C.}\ \bibnamefont {Escott}}, \bibinfo {author}
  {\bibfnamefont {A.}~\bibnamefont {Laucht}},\ and\ \bibinfo {author}
  {\bibfnamefont {A.~S.}\ \bibnamefont {Dzurak}},\ }\bibfield  {title}
  {\bibinfo {title} {Materials for silicon quantum dots and their impact on
  electron spin qubits},\ }\href {https://doi.org/10.1002/adfm.202105488}
  {\bibfield  {journal} {\bibinfo  {journal} {Advanced Functional Materials}\
  }\textbf {\bibinfo {volume} {32}},\ \bibinfo {pages} {2105488} (\bibinfo
  {year} {2022})}\BibitemShut {NoStop}%
\bibitem [{\citenamefont {Muhonen}\ \emph {et~al.}(2014)\citenamefont
  {Muhonen}, \citenamefont {Dehollain}, \citenamefont {Laucht}, \citenamefont
  {Hudson}, \citenamefont {Kalra}, \citenamefont {Sekiguchi}, \citenamefont
  {Itoh}, \citenamefont {Jamieson}, \citenamefont {McCallum}, \citenamefont
  {Dzurak},\ and\ \citenamefont {Morello}}]{muhonen2014storing}%
  \BibitemOpen
  \bibfield  {author} {\bibinfo {author} {\bibfnamefont {J.~T.}\ \bibnamefont
  {Muhonen}}, \bibinfo {author} {\bibfnamefont {J.~P.}\ \bibnamefont
  {Dehollain}}, \bibinfo {author} {\bibfnamefont {A.}~\bibnamefont {Laucht}},
  \bibinfo {author} {\bibfnamefont {F.~E.}\ \bibnamefont {Hudson}}, \bibinfo
  {author} {\bibfnamefont {R.}~\bibnamefont {Kalra}}, \bibinfo {author}
  {\bibfnamefont {T.}~\bibnamefont {Sekiguchi}}, \bibinfo {author}
  {\bibfnamefont {K.~M.}\ \bibnamefont {Itoh}}, \bibinfo {author}
  {\bibfnamefont {D.~N.}\ \bibnamefont {Jamieson}}, \bibinfo {author}
  {\bibfnamefont {J.~C.}\ \bibnamefont {McCallum}}, \bibinfo {author}
  {\bibfnamefont {A.~S.}\ \bibnamefont {Dzurak}},\ and\ \bibinfo {author}
  {\bibfnamefont {A.}~\bibnamefont {Morello}},\ }\bibfield  {title} {\bibinfo
  {title} {Storing quantum information for 30 seconds in a nanoelectronic
  device},\ }\href {https://doi.org/10.1038/nnano.2014.211} {\bibfield
  {journal} {\bibinfo  {journal} {Nature Nanotechnology}\ }\textbf {\bibinfo
  {volume} {9}},\ \bibinfo {pages} {986} (\bibinfo {year} {2014})}\BibitemShut
  {NoStop}%
\bibitem [{\citenamefont {Veldhorst}\ \emph {et~al.}(2014)\citenamefont
  {Veldhorst}, \citenamefont {Hwang}, \citenamefont {Yang}, \citenamefont
  {Leenstra}, \citenamefont {{de Ronde}}, \citenamefont {Dehollain},
  \citenamefont {Muhonen}, \citenamefont {Hudson}, \citenamefont {Itoh},
  \citenamefont {Morello},\ and\ \citenamefont
  {Dzurak}}]{veldhorst2014addressable}%
  \BibitemOpen
  \bibfield  {author} {\bibinfo {author} {\bibfnamefont {M.}~\bibnamefont
  {Veldhorst}}, \bibinfo {author} {\bibfnamefont {J.~C.~C.}\ \bibnamefont
  {Hwang}}, \bibinfo {author} {\bibfnamefont {C.~H.}\ \bibnamefont {Yang}},
  \bibinfo {author} {\bibfnamefont {A.~W.}\ \bibnamefont {Leenstra}}, \bibinfo
  {author} {\bibfnamefont {B.}~\bibnamefont {{de Ronde}}}, \bibinfo {author}
  {\bibfnamefont {J.~P.}\ \bibnamefont {Dehollain}}, \bibinfo {author}
  {\bibfnamefont {J.~T.}\ \bibnamefont {Muhonen}}, \bibinfo {author}
  {\bibfnamefont {F.~E.}\ \bibnamefont {Hudson}}, \bibinfo {author}
  {\bibfnamefont {K.~M.}\ \bibnamefont {Itoh}}, \bibinfo {author}
  {\bibfnamefont {A.}~\bibnamefont {Morello}},\ and\ \bibinfo {author}
  {\bibfnamefont {A.~S.}\ \bibnamefont {Dzurak}},\ }\bibfield  {title}
  {\bibinfo {title} {An addressable quantum dot qubit with fault-tolerant
  control-fidelity},\ }\href {https://doi.org/10.1038/nnano.2014.216}
  {\bibfield  {journal} {\bibinfo  {journal} {Nature Nanotechnology}\ }\textbf
  {\bibinfo {volume} {9}},\ \bibinfo {pages} {981} (\bibinfo {year}
  {2014})}\BibitemShut {NoStop}%
\bibitem [{\citenamefont {Takeda}\ \emph {et~al.}(2016)\citenamefont {Takeda},
  \citenamefont {Kamioka}, \citenamefont {Otsuka}, \citenamefont {Yoneda},
  \citenamefont {Nakajima}, \citenamefont {Delbecq}, \citenamefont {Amaha},
  \citenamefont {Allison}, \citenamefont {Kodera}, \citenamefont {Oda},\ and\
  \citenamefont {Tarucha}}]{takeda2016faulttolerant}%
  \BibitemOpen
  \bibfield  {author} {\bibinfo {author} {\bibfnamefont {K.}~\bibnamefont
  {Takeda}}, \bibinfo {author} {\bibfnamefont {J.}~\bibnamefont {Kamioka}},
  \bibinfo {author} {\bibfnamefont {T.}~\bibnamefont {Otsuka}}, \bibinfo
  {author} {\bibfnamefont {J.}~\bibnamefont {Yoneda}}, \bibinfo {author}
  {\bibfnamefont {T.}~\bibnamefont {Nakajima}}, \bibinfo {author}
  {\bibfnamefont {M.~R.}\ \bibnamefont {Delbecq}}, \bibinfo {author}
  {\bibfnamefont {S.}~\bibnamefont {Amaha}}, \bibinfo {author} {\bibfnamefont
  {G.}~\bibnamefont {Allison}}, \bibinfo {author} {\bibfnamefont
  {T.}~\bibnamefont {Kodera}}, \bibinfo {author} {\bibfnamefont
  {S.}~\bibnamefont {Oda}},\ and\ \bibinfo {author} {\bibfnamefont
  {S.}~\bibnamefont {Tarucha}},\ }\bibfield  {title} {\bibinfo {title} {A
  fault-tolerant addressable spin qubit in a natural silicon quantum dot},\
  }\href {https://doi.org/10.1126/sciadv.1600694} {\bibfield  {journal}
  {\bibinfo  {journal} {Science Advances}\ }\textbf {\bibinfo {volume} {2}},\
  \bibinfo {pages} {e1600694} (\bibinfo {year} {2016})}\BibitemShut {NoStop}%
\bibitem [{\citenamefont {Yoneda}\ \emph {et~al.}(2018)\citenamefont {Yoneda},
  \citenamefont {Takeda}, \citenamefont {Otsuka}, \citenamefont {Nakajima},
  \citenamefont {Delbecq}, \citenamefont {Allison}, \citenamefont {Honda},
  \citenamefont {Kodera}, \citenamefont {Oda}, \citenamefont {Hoshi},
  \citenamefont {Usami}, \citenamefont {Itoh},\ and\ \citenamefont
  {Tarucha}}]{yoneda2018quantumdot}%
  \BibitemOpen
  \bibfield  {author} {\bibinfo {author} {\bibfnamefont {J.}~\bibnamefont
  {Yoneda}}, \bibinfo {author} {\bibfnamefont {K.}~\bibnamefont {Takeda}},
  \bibinfo {author} {\bibfnamefont {T.}~\bibnamefont {Otsuka}}, \bibinfo
  {author} {\bibfnamefont {T.}~\bibnamefont {Nakajima}}, \bibinfo {author}
  {\bibfnamefont {M.~R.}\ \bibnamefont {Delbecq}}, \bibinfo {author}
  {\bibfnamefont {G.}~\bibnamefont {Allison}}, \bibinfo {author} {\bibfnamefont
  {T.}~\bibnamefont {Honda}}, \bibinfo {author} {\bibfnamefont
  {T.}~\bibnamefont {Kodera}}, \bibinfo {author} {\bibfnamefont
  {S.}~\bibnamefont {Oda}}, \bibinfo {author} {\bibfnamefont {Y.}~\bibnamefont
  {Hoshi}}, \bibinfo {author} {\bibfnamefont {N.}~\bibnamefont {Usami}},
  \bibinfo {author} {\bibfnamefont {K.~M.}\ \bibnamefont {Itoh}},\ and\
  \bibinfo {author} {\bibfnamefont {S.}~\bibnamefont {Tarucha}},\ }\bibfield
  {title} {\bibinfo {title} {A quantum-dot spin qubit with coherence limited by
  charge noise and fidelity higher than 99.9\%},\ }\href
  {https://doi.org/10.1038/s41565-017-0014-x} {\bibfield  {journal} {\bibinfo
  {journal} {Nature Nanotechnology}\ }\textbf {\bibinfo {volume} {13}},\
  \bibinfo {pages} {102} (\bibinfo {year} {2018})}\BibitemShut {NoStop}%
\bibitem [{\citenamefont {Huang}\ \emph {et~al.}(2019)\citenamefont {Huang},
  \citenamefont {Yang}, \citenamefont {Chan}, \citenamefont {Tanttu},
  \citenamefont {Hensen}, \citenamefont {Leon}, \citenamefont {Fogarty},
  \citenamefont {Hwang}, \citenamefont {Hudson}, \citenamefont {Itoh},
  \citenamefont {Morello}, \citenamefont {Laucht},\ and\ \citenamefont
  {Dzurak}}]{huang2019fidelity}%
  \BibitemOpen
  \bibfield  {author} {\bibinfo {author} {\bibfnamefont {W.}~\bibnamefont
  {Huang}}, \bibinfo {author} {\bibfnamefont {C.~H.}\ \bibnamefont {Yang}},
  \bibinfo {author} {\bibfnamefont {K.~W.}\ \bibnamefont {Chan}}, \bibinfo
  {author} {\bibfnamefont {T.}~\bibnamefont {Tanttu}}, \bibinfo {author}
  {\bibfnamefont {B.}~\bibnamefont {Hensen}}, \bibinfo {author} {\bibfnamefont
  {R.~C.~C.}\ \bibnamefont {Leon}}, \bibinfo {author} {\bibfnamefont {M.~A.}\
  \bibnamefont {Fogarty}}, \bibinfo {author} {\bibfnamefont {J.~C.~C.}\
  \bibnamefont {Hwang}}, \bibinfo {author} {\bibfnamefont {F.~E.}\ \bibnamefont
  {Hudson}}, \bibinfo {author} {\bibfnamefont {K.~M.}\ \bibnamefont {Itoh}},
  \bibinfo {author} {\bibfnamefont {A.}~\bibnamefont {Morello}}, \bibinfo
  {author} {\bibfnamefont {A.}~\bibnamefont {Laucht}},\ and\ \bibinfo {author}
  {\bibfnamefont {A.~S.}\ \bibnamefont {Dzurak}},\ }\bibfield  {title}
  {\bibinfo {title} {Fidelity benchmarks for two-qubit gates in silicon},\
  }\href {https://doi.org/10.1038/s41586-019-1197-0} {\bibfield  {journal}
  {\bibinfo  {journal} {Nature}\ }\textbf {\bibinfo {volume} {569}},\ \bibinfo
  {pages} {532} (\bibinfo {year} {2019})}\BibitemShut {NoStop}%
\bibitem [{\citenamefont {Xue}\ \emph {et~al.}(2019)\citenamefont {Xue},
  \citenamefont {Watson}, \citenamefont {Helsen}, \citenamefont {Ward},
  \citenamefont {Savage}, \citenamefont {Lagally}, \citenamefont {Coppersmith},
  \citenamefont {Eriksson}, \citenamefont {Wehner},\ and\ \citenamefont
  {Vandersypen}}]{xue2019benchmarking}%
  \BibitemOpen
  \bibfield  {author} {\bibinfo {author} {\bibfnamefont {X.}~\bibnamefont
  {Xue}}, \bibinfo {author} {\bibfnamefont {T.~F.}\ \bibnamefont {Watson}},
  \bibinfo {author} {\bibfnamefont {J.}~\bibnamefont {Helsen}}, \bibinfo
  {author} {\bibfnamefont {D.~R.}\ \bibnamefont {Ward}}, \bibinfo {author}
  {\bibfnamefont {D.~E.}\ \bibnamefont {Savage}}, \bibinfo {author}
  {\bibfnamefont {M.~G.}\ \bibnamefont {Lagally}}, \bibinfo {author}
  {\bibfnamefont {S.~N.}\ \bibnamefont {Coppersmith}}, \bibinfo {author}
  {\bibfnamefont {M.~A.}\ \bibnamefont {Eriksson}}, \bibinfo {author}
  {\bibfnamefont {S.}~\bibnamefont {Wehner}},\ and\ \bibinfo {author}
  {\bibfnamefont {L.~M.~K.}\ \bibnamefont {Vandersypen}},\ }\bibfield  {title}
  {\bibinfo {title} {Benchmarking gate fidelities in a
  $\mathrm{Si}/\mathrm{SiGe}$ two-qubit device},\ }\href
  {https://doi.org/10.1103/PhysRevX.9.021011} {\bibfield  {journal} {\bibinfo
  {journal} {Physical Review X}\ }\textbf {\bibinfo {volume} {9}},\ \bibinfo
  {pages} {021011} (\bibinfo {year} {2019})}\BibitemShut {NoStop}%
\bibitem [{\citenamefont {M{\k{a}}dzik}\ \emph {et~al.}(2022)\citenamefont
  {M{\k{a}}dzik}, \citenamefont {Asaad}, \citenamefont {Youssry}, \citenamefont
  {Joecker}, \citenamefont {Rudinger}, \citenamefont {Nielsen}, \citenamefont
  {Young}, \citenamefont {Proctor}, \citenamefont {Baczewski}, \citenamefont
  {Laucht}, \citenamefont {Schmitt}, \citenamefont {Hudson}, \citenamefont
  {Itoh}, \citenamefont {Jakob}, \citenamefont {Johnson}, \citenamefont
  {Jamieson}, \citenamefont {Dzurak}, \citenamefont {Ferrie}, \citenamefont
  {{Blume-Kohout}},\ and\ \citenamefont {Morello}}]{madzik2022precision}%
  \BibitemOpen
  \bibfield  {author} {\bibinfo {author} {\bibfnamefont {M.~T.}\ \bibnamefont
  {M{\k{a}}dzik}}, \bibinfo {author} {\bibfnamefont {S.}~\bibnamefont {Asaad}},
  \bibinfo {author} {\bibfnamefont {A.}~\bibnamefont {Youssry}}, \bibinfo
  {author} {\bibfnamefont {B.}~\bibnamefont {Joecker}}, \bibinfo {author}
  {\bibfnamefont {K.~M.}\ \bibnamefont {Rudinger}}, \bibinfo {author}
  {\bibfnamefont {E.}~\bibnamefont {Nielsen}}, \bibinfo {author} {\bibfnamefont
  {K.~C.}\ \bibnamefont {Young}}, \bibinfo {author} {\bibfnamefont {T.~J.}\
  \bibnamefont {Proctor}}, \bibinfo {author} {\bibfnamefont {A.~D.}\
  \bibnamefont {Baczewski}}, \bibinfo {author} {\bibfnamefont {A.}~\bibnamefont
  {Laucht}}, \bibinfo {author} {\bibfnamefont {V.}~\bibnamefont {Schmitt}},
  \bibinfo {author} {\bibfnamefont {F.~E.}\ \bibnamefont {Hudson}}, \bibinfo
  {author} {\bibfnamefont {K.~M.}\ \bibnamefont {Itoh}}, \bibinfo {author}
  {\bibfnamefont {A.~M.}\ \bibnamefont {Jakob}}, \bibinfo {author}
  {\bibfnamefont {B.~C.}\ \bibnamefont {Johnson}}, \bibinfo {author}
  {\bibfnamefont {D.~N.}\ \bibnamefont {Jamieson}}, \bibinfo {author}
  {\bibfnamefont {A.~S.}\ \bibnamefont {Dzurak}}, \bibinfo {author}
  {\bibfnamefont {C.}~\bibnamefont {Ferrie}}, \bibinfo {author} {\bibfnamefont
  {R.}~\bibnamefont {{Blume-Kohout}}},\ and\ \bibinfo {author} {\bibfnamefont
  {A.}~\bibnamefont {Morello}},\ }\bibfield  {title} {\bibinfo {title}
  {Precision tomography of a three-qubit donor quantum processor in silicon},\
  }\href {https://doi.org/10.1038/s41586-021-04292-7} {\bibfield  {journal}
  {\bibinfo  {journal} {Nature}\ }\textbf {\bibinfo {volume} {601}},\ \bibinfo
  {pages} {348} (\bibinfo {year} {2022})}\BibitemShut {NoStop}%
\bibitem [{\citenamefont {Xue}\ \emph {et~al.}(2022)\citenamefont {Xue},
  \citenamefont {Russ}, \citenamefont {Samkharadze}, \citenamefont {Undseth},
  \citenamefont {Sammak}, \citenamefont {Scappucci},\ and\ \citenamefont
  {Vandersypen}}]{xue2022quantum}%
  \BibitemOpen
  \bibfield  {author} {\bibinfo {author} {\bibfnamefont {X.}~\bibnamefont
  {Xue}}, \bibinfo {author} {\bibfnamefont {M.}~\bibnamefont {Russ}}, \bibinfo
  {author} {\bibfnamefont {N.}~\bibnamefont {Samkharadze}}, \bibinfo {author}
  {\bibfnamefont {B.}~\bibnamefont {Undseth}}, \bibinfo {author} {\bibfnamefont
  {A.}~\bibnamefont {Sammak}}, \bibinfo {author} {\bibfnamefont
  {G.}~\bibnamefont {Scappucci}},\ and\ \bibinfo {author} {\bibfnamefont
  {L.~M.~K.}\ \bibnamefont {Vandersypen}},\ }\bibfield  {title} {\bibinfo
  {title} {Quantum logic with spin qubits crossing the surface code
  threshold},\ }\href {https://doi.org/10.1038/s41586-021-04273-w} {\bibfield
  {journal} {\bibinfo  {journal} {Nature}\ }\textbf {\bibinfo {volume} {601}},\
  \bibinfo {pages} {343} (\bibinfo {year} {2022})}\BibitemShut {NoStop}%
\bibitem [{\citenamefont {Noiri}\ \emph
  {et~al.}(2022{\natexlab{a}})\citenamefont {Noiri}, \citenamefont {Takeda},
  \citenamefont {Nakajima}, \citenamefont {Kobayashi}, \citenamefont {Sammak},
  \citenamefont {Scappucci},\ and\ \citenamefont {Tarucha}}]{noiri2022fast}%
  \BibitemOpen
  \bibfield  {author} {\bibinfo {author} {\bibfnamefont {A.}~\bibnamefont
  {Noiri}}, \bibinfo {author} {\bibfnamefont {K.}~\bibnamefont {Takeda}},
  \bibinfo {author} {\bibfnamefont {T.}~\bibnamefont {Nakajima}}, \bibinfo
  {author} {\bibfnamefont {T.}~\bibnamefont {Kobayashi}}, \bibinfo {author}
  {\bibfnamefont {A.}~\bibnamefont {Sammak}}, \bibinfo {author} {\bibfnamefont
  {G.}~\bibnamefont {Scappucci}},\ and\ \bibinfo {author} {\bibfnamefont
  {S.}~\bibnamefont {Tarucha}},\ }\bibfield  {title} {\bibinfo {title} {Fast
  universal quantum gate above the fault-tolerance threshold in silicon},\
  }\href {https://doi.org/10.1038/s41586-021-04182-y} {\bibfield  {journal}
  {\bibinfo  {journal} {Nature}\ }\textbf {\bibinfo {volume} {601}},\ \bibinfo
  {pages} {338} (\bibinfo {year} {2022}{\natexlab{a}})}\BibitemShut {NoStop}%
\bibitem [{\citenamefont {Mills}\ \emph {et~al.}(2022)\citenamefont {Mills},
  \citenamefont {Guinn}, \citenamefont {Feldman}, \citenamefont {Sigillito},
  \citenamefont {Gullans}, \citenamefont {Rakher}, \citenamefont {Kerckhoff},
  \citenamefont {Jackson},\ and\ \citenamefont {Petta}}]{mills2022high}%
  \BibitemOpen
  \bibfield  {author} {\bibinfo {author} {\bibfnamefont {A.~R.}\ \bibnamefont
  {Mills}}, \bibinfo {author} {\bibfnamefont {C.~R.}\ \bibnamefont {Guinn}},
  \bibinfo {author} {\bibfnamefont {M.~M.}\ \bibnamefont {Feldman}}, \bibinfo
  {author} {\bibfnamefont {A.~J.}\ \bibnamefont {Sigillito}}, \bibinfo {author}
  {\bibfnamefont {M.~J.}\ \bibnamefont {Gullans}}, \bibinfo {author}
  {\bibfnamefont {M.}~\bibnamefont {Rakher}}, \bibinfo {author} {\bibfnamefont
  {J.}~\bibnamefont {Kerckhoff}}, \bibinfo {author} {\bibfnamefont {C.~A.~C.}\
  \bibnamefont {Jackson}},\ and\ \bibinfo {author} {\bibfnamefont {J.~R.}\
  \bibnamefont {Petta}},\ }\href {https://doi.org/10.48550/arXiv.2204.09551}
  {\bibinfo {title} {High fidelity state preparation, quantum control, and
  readout of an isotopically enriched silicon spin qubit}} (\bibinfo {year}
  {2022}),\ \Eprint {https://arxiv.org/abs/2204.09551} {arXiv:2204.09551
  [cond-mat, physics:quant-ph]} \BibitemShut {NoStop}%
\bibitem [{\citenamefont {Mehl}\ \emph {et~al.}(2014)\citenamefont {Mehl},
  \citenamefont {Bluhm},\ and\ \citenamefont {DiVincenzo}}]{mehl2014twoqubit}%
  \BibitemOpen
  \bibfield  {author} {\bibinfo {author} {\bibfnamefont {S.}~\bibnamefont
  {Mehl}}, \bibinfo {author} {\bibfnamefont {H.}~\bibnamefont {Bluhm}},\ and\
  \bibinfo {author} {\bibfnamefont {D.~P.}\ \bibnamefont {DiVincenzo}},\
  }\bibfield  {title} {\bibinfo {title} {Two-qubit couplings of singlet-triplet
  qubits mediated by one quantum state},\ }\href
  {https://doi.org/10.1103/PhysRevB.90.045404} {\bibfield  {journal} {\bibinfo
  {journal} {Physical Review B}\ }\textbf {\bibinfo {volume} {90}},\ \bibinfo
  {pages} {045404} (\bibinfo {year} {2014})}\BibitemShut {NoStop}%
\bibitem [{\citenamefont {Srinivasa}\ \emph {et~al.}(2015)\citenamefont
  {Srinivasa}, \citenamefont {Xu},\ and\ \citenamefont
  {Taylor}}]{srinivasa2015tunable}%
  \BibitemOpen
  \bibfield  {author} {\bibinfo {author} {\bibfnamefont {V.}~\bibnamefont
  {Srinivasa}}, \bibinfo {author} {\bibfnamefont {H.}~\bibnamefont {Xu}},\ and\
  \bibinfo {author} {\bibfnamefont {J.~M.}\ \bibnamefont {Taylor}},\ }\bibfield
   {title} {\bibinfo {title} {Tunable spin-qubit coupling mediated by a
  multielectron quantum dot},\ }\href
  {https://doi.org/10.1103/PhysRevLett.114.226803} {\bibfield  {journal}
  {\bibinfo  {journal} {Physical Review Letters}\ }\textbf {\bibinfo {volume}
  {114}},\ \bibinfo {pages} {226803} (\bibinfo {year} {2015})}\BibitemShut
  {NoStop}%
\bibitem [{\citenamefont {Stano}\ \emph {et~al.}(2015)\citenamefont {Stano},
  \citenamefont {Klinovaja}, \citenamefont {Braakman}, \citenamefont
  {Vandersypen},\ and\ \citenamefont {Loss}}]{stano2015fast}%
  \BibitemOpen
  \bibfield  {author} {\bibinfo {author} {\bibfnamefont {P.}~\bibnamefont
  {Stano}}, \bibinfo {author} {\bibfnamefont {J.}~\bibnamefont {Klinovaja}},
  \bibinfo {author} {\bibfnamefont {F.~R.}\ \bibnamefont {Braakman}}, \bibinfo
  {author} {\bibfnamefont {L.~M.~K.}\ \bibnamefont {Vandersypen}},\ and\
  \bibinfo {author} {\bibfnamefont {D.}~\bibnamefont {Loss}},\ }\bibfield
  {title} {\bibinfo {title} {Fast long-distance control of spin qubits by
  photon-assisted cotunneling},\ }\href
  {https://doi.org/10.1103/PhysRevB.92.075302} {\bibfield  {journal} {\bibinfo
  {journal} {Physical Review B}\ }\textbf {\bibinfo {volume} {92}},\ \bibinfo
  {pages} {075302} (\bibinfo {year} {2015})}\BibitemShut {NoStop}%
\bibitem [{\citenamefont {Mohiyaddin}\ \emph {et~al.}(2016)\citenamefont
  {Mohiyaddin}, \citenamefont {Kalra}, \citenamefont {Laucht}, \citenamefont
  {Rahman}, \citenamefont {Klimeck},\ and\ \citenamefont
  {Morello}}]{mohiyaddin2016transport}%
  \BibitemOpen
  \bibfield  {author} {\bibinfo {author} {\bibfnamefont {F.~A.}\ \bibnamefont
  {Mohiyaddin}}, \bibinfo {author} {\bibfnamefont {R.}~\bibnamefont {Kalra}},
  \bibinfo {author} {\bibfnamefont {A.}~\bibnamefont {Laucht}}, \bibinfo
  {author} {\bibfnamefont {R.}~\bibnamefont {Rahman}}, \bibinfo {author}
  {\bibfnamefont {G.}~\bibnamefont {Klimeck}},\ and\ \bibinfo {author}
  {\bibfnamefont {A.}~\bibnamefont {Morello}},\ }\bibfield  {title} {\bibinfo
  {title} {Transport of spin qubits with donor chains under realistic
  experimental conditions},\ }\href
  {https://doi.org/10.1103/PhysRevB.94.045314} {\bibfield  {journal} {\bibinfo
  {journal} {Physical Review B}\ }\textbf {\bibinfo {volume} {94}},\ \bibinfo
  {pages} {045314} (\bibinfo {year} {2016})}\BibitemShut {NoStop}%
\bibitem [{\citenamefont {Vandersypen}\ \emph {et~al.}(2017)\citenamefont
  {Vandersypen}, \citenamefont {Bluhm}, \citenamefont {Clarke}, \citenamefont
  {Dzurak}, \citenamefont {Ishihara}, \citenamefont {Morello}, \citenamefont
  {Reilly}, \citenamefont {Schreiber},\ and\ \citenamefont
  {Veldhorst}}]{vandersypen2017interfacing}%
  \BibitemOpen
  \bibfield  {author} {\bibinfo {author} {\bibfnamefont {L.~M.~K.}\
  \bibnamefont {Vandersypen}}, \bibinfo {author} {\bibfnamefont
  {H.}~\bibnamefont {Bluhm}}, \bibinfo {author} {\bibfnamefont {J.~S.}\
  \bibnamefont {Clarke}}, \bibinfo {author} {\bibfnamefont {A.~S.}\
  \bibnamefont {Dzurak}}, \bibinfo {author} {\bibfnamefont {R.}~\bibnamefont
  {Ishihara}}, \bibinfo {author} {\bibfnamefont {A.}~\bibnamefont {Morello}},
  \bibinfo {author} {\bibfnamefont {D.~J.}\ \bibnamefont {Reilly}}, \bibinfo
  {author} {\bibfnamefont {L.~R.}\ \bibnamefont {Schreiber}},\ and\ \bibinfo
  {author} {\bibfnamefont {M.}~\bibnamefont {Veldhorst}},\ }\bibfield  {title}
  {\bibinfo {title} {Interfacing spin qubits in quantum dots and
  donors\textemdash hot, dense, and coherent},\ }\href
  {https://doi.org/10.1038/s41534-017-0038-y} {\bibfield  {journal} {\bibinfo
  {journal} {npj Quantum Information}\ }\textbf {\bibinfo {volume} {3}},\
  \bibinfo {pages} {1} (\bibinfo {year} {2017})}\BibitemShut {NoStop}%
\bibitem [{\citenamefont {Baart}\ \emph {et~al.}(2016)\citenamefont {Baart},
  \citenamefont {Shafiei}, \citenamefont {Fujita}, \citenamefont {Reichl},
  \citenamefont {Wegscheider},\ and\ \citenamefont
  {Vandersypen}}]{baart2016singlespin}%
  \BibitemOpen
  \bibfield  {author} {\bibinfo {author} {\bibfnamefont {T.~A.}\ \bibnamefont
  {Baart}}, \bibinfo {author} {\bibfnamefont {M.}~\bibnamefont {Shafiei}},
  \bibinfo {author} {\bibfnamefont {T.}~\bibnamefont {Fujita}}, \bibinfo
  {author} {\bibfnamefont {C.}~\bibnamefont {Reichl}}, \bibinfo {author}
  {\bibfnamefont {W.}~\bibnamefont {Wegscheider}},\ and\ \bibinfo {author}
  {\bibfnamefont {L.~M.~K.}\ \bibnamefont {Vandersypen}},\ }\bibfield  {title}
  {\bibinfo {title} {Single-spin {CCD}},\ }\href
  {https://doi.org/10.1038/nnano.2015.291} {\bibfield  {journal} {\bibinfo
  {journal} {Nature Nanotechnology}\ }\textbf {\bibinfo {volume} {11}},\
  \bibinfo {pages} {330} (\bibinfo {year} {2016})}\BibitemShut {NoStop}%
\bibitem [{\citenamefont {Fujita}\ \emph {et~al.}(2017)\citenamefont {Fujita},
  \citenamefont {Baart}, \citenamefont {Reichl}, \citenamefont {Wegscheider},\
  and\ \citenamefont {Vandersypen}}]{fujita2017coherent}%
  \BibitemOpen
  \bibfield  {author} {\bibinfo {author} {\bibfnamefont {T.}~\bibnamefont
  {Fujita}}, \bibinfo {author} {\bibfnamefont {T.~A.}\ \bibnamefont {Baart}},
  \bibinfo {author} {\bibfnamefont {C.}~\bibnamefont {Reichl}}, \bibinfo
  {author} {\bibfnamefont {W.}~\bibnamefont {Wegscheider}},\ and\ \bibinfo
  {author} {\bibfnamefont {L.~M.~K.}\ \bibnamefont {Vandersypen}},\ }\bibfield
  {title} {\bibinfo {title} {Coherent shuttle of electron-spin states},\ }\href
  {https://doi.org/10.1038/s41534-017-0024-4} {\bibfield  {journal} {\bibinfo
  {journal} {npj Quantum Information}\ }\textbf {\bibinfo {volume} {3}},\
  \bibinfo {pages} {1} (\bibinfo {year} {2017})}\BibitemShut {NoStop}%
\bibitem [{\citenamefont {Fedele}\ \emph {et~al.}(2021)\citenamefont {Fedele},
  \citenamefont {Chatterjee}, \citenamefont {Fallahi}, \citenamefont {Gardner},
  \citenamefont {Manfra},\ and\ \citenamefont
  {Kuemmeth}}]{fedele2021simultaneous}%
  \BibitemOpen
  \bibfield  {author} {\bibinfo {author} {\bibfnamefont {F.}~\bibnamefont
  {Fedele}}, \bibinfo {author} {\bibfnamefont {A.}~\bibnamefont {Chatterjee}},
  \bibinfo {author} {\bibfnamefont {S.}~\bibnamefont {Fallahi}}, \bibinfo
  {author} {\bibfnamefont {G.~C.}\ \bibnamefont {Gardner}}, \bibinfo {author}
  {\bibfnamefont {M.~J.}\ \bibnamefont {Manfra}},\ and\ \bibinfo {author}
  {\bibfnamefont {F.}~\bibnamefont {Kuemmeth}},\ }\bibfield  {title} {\bibinfo
  {title} {Simultaneous operations in a two-dimensional array of
  singlet-triplet qubits},\ }\href
  {https://doi.org/10.1103/PRXQuantum.2.040306} {\bibfield  {journal} {\bibinfo
   {journal} {PRX Quantum}\ }\textbf {\bibinfo {volume} {2}},\ \bibinfo {pages}
  {040306} (\bibinfo {year} {2021})}\BibitemShut {NoStop}%
\bibitem [{\citenamefont {Sigillito}\ \emph {et~al.}(2019)\citenamefont
  {Sigillito}, \citenamefont {Loy}, \citenamefont {Zajac}, \citenamefont
  {Gullans}, \citenamefont {Edge},\ and\ \citenamefont
  {Petta}}]{sigillito2019siteselective}%
  \BibitemOpen
  \bibfield  {author} {\bibinfo {author} {\bibfnamefont {A.}~\bibnamefont
  {Sigillito}}, \bibinfo {author} {\bibfnamefont {J.}~\bibnamefont {Loy}},
  \bibinfo {author} {\bibfnamefont {D.}~\bibnamefont {Zajac}}, \bibinfo
  {author} {\bibfnamefont {M.}~\bibnamefont {Gullans}}, \bibinfo {author}
  {\bibfnamefont {L.}~\bibnamefont {Edge}},\ and\ \bibinfo {author}
  {\bibfnamefont {J.}~\bibnamefont {Petta}},\ }\bibfield  {title} {\bibinfo
  {title} {Site-selective quantum control in an isotopically enriched
  $^{28}\mathrm{Si}/\mathrm{Si}_{0.7}\mathrm{Ge}_{0.3}$ quadruple quantum
  dot},\ }\href {https://doi.org/10.1103/PhysRevApplied.11.061006} {\bibfield
  {journal} {\bibinfo  {journal} {Physical Review Applied}\ }\textbf {\bibinfo
  {volume} {11}},\ \bibinfo {pages} {061006} (\bibinfo {year}
  {2019})}\BibitemShut {NoStop}%
\bibitem [{\citenamefont {Yoneda}\ \emph {et~al.}(2021)\citenamefont {Yoneda},
  \citenamefont {Huang}, \citenamefont {Feng}, \citenamefont {Yang},
  \citenamefont {Chan}, \citenamefont {Tanttu}, \citenamefont {Gilbert},
  \citenamefont {Leon}, \citenamefont {Hudson}, \citenamefont {Itoh},
  \citenamefont {Morello}, \citenamefont {Bartlett}, \citenamefont {Laucht},
  \citenamefont {Saraiva},\ and\ \citenamefont {Dzurak}}]{yoneda2021coherent}%
  \BibitemOpen
  \bibfield  {author} {\bibinfo {author} {\bibfnamefont {J.}~\bibnamefont
  {Yoneda}}, \bibinfo {author} {\bibfnamefont {W.}~\bibnamefont {Huang}},
  \bibinfo {author} {\bibfnamefont {M.}~\bibnamefont {Feng}}, \bibinfo {author}
  {\bibfnamefont {C.~H.}\ \bibnamefont {Yang}}, \bibinfo {author}
  {\bibfnamefont {K.~W.}\ \bibnamefont {Chan}}, \bibinfo {author}
  {\bibfnamefont {T.}~\bibnamefont {Tanttu}}, \bibinfo {author} {\bibfnamefont
  {W.}~\bibnamefont {Gilbert}}, \bibinfo {author} {\bibfnamefont {R.~C.~C.}\
  \bibnamefont {Leon}}, \bibinfo {author} {\bibfnamefont {F.~E.}\ \bibnamefont
  {Hudson}}, \bibinfo {author} {\bibfnamefont {K.~M.}\ \bibnamefont {Itoh}},
  \bibinfo {author} {\bibfnamefont {A.}~\bibnamefont {Morello}}, \bibinfo
  {author} {\bibfnamefont {S.~D.}\ \bibnamefont {Bartlett}}, \bibinfo {author}
  {\bibfnamefont {A.}~\bibnamefont {Laucht}}, \bibinfo {author} {\bibfnamefont
  {A.}~\bibnamefont {Saraiva}},\ and\ \bibinfo {author} {\bibfnamefont {A.~S.}\
  \bibnamefont {Dzurak}},\ }\bibfield  {title} {\bibinfo {title} {Coherent spin
  qubit transport in silicon},\ }\href
  {https://doi.org/10.1038/s41467-021-24371-7} {\bibfield  {journal} {\bibinfo
  {journal} {Nature Communications}\ }\textbf {\bibinfo {volume} {12}},\
  \bibinfo {pages} {4114} (\bibinfo {year} {2021})}\BibitemShut {NoStop}%
\bibitem [{\citenamefont {Noiri}\ \emph
  {et~al.}(2022{\natexlab{b}})\citenamefont {Noiri}, \citenamefont {Takeda},
  \citenamefont {Nakajima}, \citenamefont {Kobayashi}, \citenamefont {Sammak},
  \citenamefont {Scappucci},\ and\ \citenamefont
  {Tarucha}}]{noiri2022shuttlingbased}%
  \BibitemOpen
  \bibfield  {author} {\bibinfo {author} {\bibfnamefont {A.}~\bibnamefont
  {Noiri}}, \bibinfo {author} {\bibfnamefont {K.}~\bibnamefont {Takeda}},
  \bibinfo {author} {\bibfnamefont {T.}~\bibnamefont {Nakajima}}, \bibinfo
  {author} {\bibfnamefont {T.}~\bibnamefont {Kobayashi}}, \bibinfo {author}
  {\bibfnamefont {A.}~\bibnamefont {Sammak}}, \bibinfo {author} {\bibfnamefont
  {G.}~\bibnamefont {Scappucci}},\ and\ \bibinfo {author} {\bibfnamefont
  {S.}~\bibnamefont {Tarucha}},\ }\href
  {https://doi.org/10.48550/arXiv.2202.01357} {\bibinfo {title} {A
  shuttling-based two-qubit logic gate for linking distant silicon quantum
  processors}} (\bibinfo {year} {2022}{\natexlab{b}}),\ \Eprint
  {https://arxiv.org/abs/2202.01357} {arXiv:2202.01357 [cond-mat,
  physics:quant-ph]} \BibitemShut {NoStop}%
\bibitem [{\citenamefont {Croot}\ \emph {et~al.}(2018)\citenamefont {Croot},
  \citenamefont {Pauka}, \citenamefont {Watson}, \citenamefont {Gardner},
  \citenamefont {Fallahi}, \citenamefont {Manfra},\ and\ \citenamefont
  {Reilly}}]{croot2018device}%
  \BibitemOpen
  \bibfield  {author} {\bibinfo {author} {\bibfnamefont {X.}~\bibnamefont
  {Croot}}, \bibinfo {author} {\bibfnamefont {S.}~\bibnamefont {Pauka}},
  \bibinfo {author} {\bibfnamefont {J.}~\bibnamefont {Watson}}, \bibinfo
  {author} {\bibfnamefont {G.}~\bibnamefont {Gardner}}, \bibinfo {author}
  {\bibfnamefont {S.}~\bibnamefont {Fallahi}}, \bibinfo {author} {\bibfnamefont
  {M.}~\bibnamefont {Manfra}},\ and\ \bibinfo {author} {\bibfnamefont
  {D.}~\bibnamefont {Reilly}},\ }\bibfield  {title} {\bibinfo {title} {Device
  architecture for coupling spin qubits via an intermediate quantum state},\
  }\href {https://doi.org/10.1103/PhysRevApplied.10.044058} {\bibfield
  {journal} {\bibinfo  {journal} {Physical Review Applied}\ }\textbf {\bibinfo
  {volume} {10}},\ \bibinfo {pages} {044058} (\bibinfo {year}
  {2018})}\BibitemShut {NoStop}%
\bibitem [{\citenamefont {Malinowski}\ \emph {et~al.}(2018)\citenamefont
  {Malinowski}, \citenamefont {Martins}, \citenamefont {Smith}, \citenamefont
  {Bartlett}, \citenamefont {Doherty}, \citenamefont {Nissen}, \citenamefont
  {Fallahi}, \citenamefont {Gardner}, \citenamefont {Manfra}, \citenamefont
  {Marcus},\ and\ \citenamefont {Kuemmeth}}]{malinowski2018spin}%
  \BibitemOpen
  \bibfield  {author} {\bibinfo {author} {\bibfnamefont {F.~K.}\ \bibnamefont
  {Malinowski}}, \bibinfo {author} {\bibfnamefont {F.}~\bibnamefont {Martins}},
  \bibinfo {author} {\bibfnamefont {T.~B.}\ \bibnamefont {Smith}}, \bibinfo
  {author} {\bibfnamefont {S.~D.}\ \bibnamefont {Bartlett}}, \bibinfo {author}
  {\bibfnamefont {A.~C.}\ \bibnamefont {Doherty}}, \bibinfo {author}
  {\bibfnamefont {P.~D.}\ \bibnamefont {Nissen}}, \bibinfo {author}
  {\bibfnamefont {S.}~\bibnamefont {Fallahi}}, \bibinfo {author} {\bibfnamefont
  {G.~C.}\ \bibnamefont {Gardner}}, \bibinfo {author} {\bibfnamefont {M.~J.}\
  \bibnamefont {Manfra}}, \bibinfo {author} {\bibfnamefont {C.~M.}\
  \bibnamefont {Marcus}},\ and\ \bibinfo {author} {\bibfnamefont
  {F.}~\bibnamefont {Kuemmeth}},\ }\bibfield  {title} {\bibinfo {title} {Spin
  of a multielectron quantum dot and its interaction with a neighboring
  electron},\ }\href {https://doi.org/10.1103/PhysRevX.8.011045} {\bibfield
  {journal} {\bibinfo  {journal} {Physical Review X}\ }\textbf {\bibinfo
  {volume} {8}},\ \bibinfo {pages} {011045} (\bibinfo {year}
  {2018})}\BibitemShut {NoStop}%
\bibitem [{\citenamefont {Malinowski}\ \emph {et~al.}(2019)\citenamefont
  {Malinowski}, \citenamefont {Martins}, \citenamefont {Smith}, \citenamefont
  {Bartlett}, \citenamefont {Doherty}, \citenamefont {Nissen}, \citenamefont
  {Fallahi}, \citenamefont {Gardner}, \citenamefont {Manfra}, \citenamefont
  {Marcus},\ and\ \citenamefont {Kuemmeth}}]{malinowski2019fast}%
  \BibitemOpen
  \bibfield  {author} {\bibinfo {author} {\bibfnamefont {F.~K.}\ \bibnamefont
  {Malinowski}}, \bibinfo {author} {\bibfnamefont {F.}~\bibnamefont {Martins}},
  \bibinfo {author} {\bibfnamefont {T.~B.}\ \bibnamefont {Smith}}, \bibinfo
  {author} {\bibfnamefont {S.~D.}\ \bibnamefont {Bartlett}}, \bibinfo {author}
  {\bibfnamefont {A.~C.}\ \bibnamefont {Doherty}}, \bibinfo {author}
  {\bibfnamefont {P.~D.}\ \bibnamefont {Nissen}}, \bibinfo {author}
  {\bibfnamefont {S.}~\bibnamefont {Fallahi}}, \bibinfo {author} {\bibfnamefont
  {G.~C.}\ \bibnamefont {Gardner}}, \bibinfo {author} {\bibfnamefont {M.~J.}\
  \bibnamefont {Manfra}}, \bibinfo {author} {\bibfnamefont {C.~M.}\
  \bibnamefont {Marcus}},\ and\ \bibinfo {author} {\bibfnamefont
  {F.}~\bibnamefont {Kuemmeth}},\ }\bibfield  {title} {\bibinfo {title} {Fast
  spin exchange across a multielectron mediator},\ }\href
  {https://doi.org/10.1038/s41467-019-09194-x} {\bibfield  {journal} {\bibinfo
  {journal} {Nature Communications}\ }\textbf {\bibinfo {volume} {10}},\
  \bibinfo {pages} {1196} (\bibinfo {year} {2019})}\BibitemShut {NoStop}%
\bibitem [{\citenamefont {Yannouleas}\ and\ \citenamefont
  {Landman}(1999)}]{yannouleas1999spontaneous}%
  \BibitemOpen
  \bibfield  {author} {\bibinfo {author} {\bibfnamefont {C.}~\bibnamefont
  {Yannouleas}}\ and\ \bibinfo {author} {\bibfnamefont {U.}~\bibnamefont
  {Landman}},\ }\bibfield  {title} {\bibinfo {title} {Spontaneous symmetry
  breaking in single and molecular quantum dots},\ }\href
  {https://doi.org/10.1103/PhysRevLett.82.5325} {\bibfield  {journal} {\bibinfo
   {journal} {Physical Review Letters}\ }\textbf {\bibinfo {volume} {82}},\
  \bibinfo {pages} {5325} (\bibinfo {year} {1999})}\BibitemShut {NoStop}%
\bibitem [{\citenamefont {Cavaliere}\ \emph {et~al.}(2009)\citenamefont
  {Cavaliere}, \citenamefont {Giovannini}, \citenamefont {Sassetti},\ and\
  \citenamefont {Kramer}}]{cavaliere2009transport}%
  \BibitemOpen
  \bibfield  {author} {\bibinfo {author} {\bibfnamefont {F.}~\bibnamefont
  {Cavaliere}}, \bibinfo {author} {\bibfnamefont {U.~D.}\ \bibnamefont
  {Giovannini}}, \bibinfo {author} {\bibfnamefont {M.}~\bibnamefont
  {Sassetti}},\ and\ \bibinfo {author} {\bibfnamefont {B.}~\bibnamefont
  {Kramer}},\ }\bibfield  {title} {\bibinfo {title} {Transport properties of
  quantum dots in the {Wigner} molecule regime},\ }\href
  {https://doi.org/10.1088/1367-2630/11/12/123004} {\bibfield  {journal}
  {\bibinfo  {journal} {New Journal of Physics}\ }\textbf {\bibinfo {volume}
  {11}},\ \bibinfo {pages} {123004} (\bibinfo {year} {2009})}\BibitemShut
  {NoStop}%
\bibitem [{\citenamefont {Corrigan}\ \emph {et~al.}(2021)\citenamefont
  {Corrigan}, \citenamefont {Dodson}, \citenamefont {Ercan}, \citenamefont
  {{Abadillo-Uriel}}, \citenamefont {Thorgrimsson}, \citenamefont {Knapp},
  \citenamefont {Holman}, \citenamefont {McJunkin}, \citenamefont {Neyens},
  \citenamefont {MacQuarrie}, \citenamefont {Foote}, \citenamefont {Edge},
  \citenamefont {Friesen}, \citenamefont {Coppersmith},\ and\ \citenamefont
  {Eriksson}}]{corrigan2021coherent}%
  \BibitemOpen
  \bibfield  {author} {\bibinfo {author} {\bibfnamefont {J.}~\bibnamefont
  {Corrigan}}, \bibinfo {author} {\bibfnamefont {J.~P.}\ \bibnamefont
  {Dodson}}, \bibinfo {author} {\bibfnamefont {H.~E.}\ \bibnamefont {Ercan}},
  \bibinfo {author} {\bibfnamefont {J.~C.}\ \bibnamefont {{Abadillo-Uriel}}},
  \bibinfo {author} {\bibfnamefont {B.}~\bibnamefont {Thorgrimsson}}, \bibinfo
  {author} {\bibfnamefont {T.~J.}\ \bibnamefont {Knapp}}, \bibinfo {author}
  {\bibfnamefont {N.}~\bibnamefont {Holman}}, \bibinfo {author} {\bibfnamefont
  {T.}~\bibnamefont {McJunkin}}, \bibinfo {author} {\bibfnamefont {S.~F.}\
  \bibnamefont {Neyens}}, \bibinfo {author} {\bibfnamefont {E.~R.}\
  \bibnamefont {MacQuarrie}}, \bibinfo {author} {\bibfnamefont {R.~H.}\
  \bibnamefont {Foote}}, \bibinfo {author} {\bibfnamefont {L.~F.}\ \bibnamefont
  {Edge}}, \bibinfo {author} {\bibfnamefont {M.}~\bibnamefont {Friesen}},
  \bibinfo {author} {\bibfnamefont {S.~N.}\ \bibnamefont {Coppersmith}},\ and\
  \bibinfo {author} {\bibfnamefont {M.~A.}\ \bibnamefont {Eriksson}},\
  }\bibfield  {title} {\bibinfo {title} {Coherent control and spectroscopy of a
  semiconductor quantum dot {Wigner} molecule},\ }\href
  {https://doi.org/10.1103/PhysRevLett.127.127701} {\bibfield  {journal}
  {\bibinfo  {journal} {Physical Review Letters}\ }\textbf {\bibinfo {volume}
  {127}},\ \bibinfo {pages} {127701} (\bibinfo {year} {2021})}\BibitemShut
  {NoStop}%
\bibitem [{\citenamefont {{Abadillo-Uriel}}\ \emph {et~al.}(2021)\citenamefont
  {{Abadillo-Uriel}}, \citenamefont {Martinez}, \citenamefont {Filippone},\
  and\ \citenamefont {Niquet}}]{abadillo-uriel2021twobody}%
  \BibitemOpen
  \bibfield  {author} {\bibinfo {author} {\bibfnamefont {J.~C.}\ \bibnamefont
  {{Abadillo-Uriel}}}, \bibinfo {author} {\bibfnamefont {B.}~\bibnamefont
  {Martinez}}, \bibinfo {author} {\bibfnamefont {M.}~\bibnamefont
  {Filippone}},\ and\ \bibinfo {author} {\bibfnamefont {Y.-M.}\ \bibnamefont
  {Niquet}},\ }\bibfield  {title} {\bibinfo {title} {Two-body {Wigner}
  molecularization in asymmetric quantum dot spin qubits},\ }\href
  {https://doi.org/10.1103/PhysRevB.104.195305} {\bibfield  {journal} {\bibinfo
   {journal} {Physical Review B}\ }\textbf {\bibinfo {volume} {104}},\ \bibinfo
  {pages} {195305} (\bibinfo {year} {2021})}\BibitemShut {NoStop}%
\bibitem [{\citenamefont {Yannouleas}\ and\ \citenamefont
  {Landman}(2022{\natexlab{a}})}]{yannouleas2022molecular}%
  \BibitemOpen
  \bibfield  {author} {\bibinfo {author} {\bibfnamefont {C.}~\bibnamefont
  {Yannouleas}}\ and\ \bibinfo {author} {\bibfnamefont {U.}~\bibnamefont
  {Landman}},\ }\bibfield  {title} {\bibinfo {title} {Molecular formations and
  spectra due to electron correlations in three-electron hybrid double-well
  qubits},\ }\href {https://doi.org/10.1103/PhysRevB.105.205302} {\bibfield
  {journal} {\bibinfo  {journal} {Physical Review B}\ }\textbf {\bibinfo
  {volume} {105}},\ \bibinfo {pages} {205302} (\bibinfo {year}
  {2022}{\natexlab{a}})}\BibitemShut {NoStop}%
\bibitem [{\citenamefont {Yannouleas}\ and\ \citenamefont
  {Landman}(2022{\natexlab{b}})}]{yannouleas2022wigner}%
  \BibitemOpen
  \bibfield  {author} {\bibinfo {author} {\bibfnamefont {C.}~\bibnamefont
  {Yannouleas}}\ and\ \bibinfo {author} {\bibfnamefont {U.}~\bibnamefont
  {Landman}},\ }\bibfield  {title} {\bibinfo {title} {{Wigner} molecules and
  hybrid qubits},\ }\href {https://doi.org/10.1088/1361-648X/ac5c28} {\bibfield
   {journal} {\bibinfo  {journal} {Journal of Physics: Condensed Matter}\
  }\textbf {\bibinfo {volume} {34}},\ \bibinfo {pages} {21LT01} (\bibinfo
  {year} {2022}{\natexlab{b}})}\BibitemShut {NoStop}%
\bibitem [{\citenamefont {Jang}\ \emph {et~al.}(2022)\citenamefont {Jang},
  \citenamefont {Kim}, \citenamefont {Park}, \citenamefont {Kim}, \citenamefont
  {Cho}, \citenamefont {Jang}, \citenamefont {Sim}, \citenamefont {Kang},
  \citenamefont {Jung}, \citenamefont {Umansky},\ and\ \citenamefont
  {Kim}}]{jang2022wignermolecularizationenabled}%
  \BibitemOpen
  \bibfield  {author} {\bibinfo {author} {\bibfnamefont {W.}~\bibnamefont
  {Jang}}, \bibinfo {author} {\bibfnamefont {J.}~\bibnamefont {Kim}}, \bibinfo
  {author} {\bibfnamefont {J.}~\bibnamefont {Park}}, \bibinfo {author}
  {\bibfnamefont {G.}~\bibnamefont {Kim}}, \bibinfo {author} {\bibfnamefont
  {M.-K.}\ \bibnamefont {Cho}}, \bibinfo {author} {\bibfnamefont
  {H.}~\bibnamefont {Jang}}, \bibinfo {author} {\bibfnamefont {S.}~\bibnamefont
  {Sim}}, \bibinfo {author} {\bibfnamefont {B.}~\bibnamefont {Kang}}, \bibinfo
  {author} {\bibfnamefont {H.}~\bibnamefont {Jung}}, \bibinfo {author}
  {\bibfnamefont {V.}~\bibnamefont {Umansky}},\ and\ \bibinfo {author}
  {\bibfnamefont {D.}~\bibnamefont {Kim}},\ }\href
  {https://doi.org/10.48550/arXiv.2207.11655} {\bibinfo {title}
  {{Wigner}-molecularization-enabled dynamic nuclear field programming}}
  (\bibinfo {year} {2022}),\ \Eprint {https://arxiv.org/abs/2207.11655}
  {arXiv:2207.11655 [cond-mat, physics:quant-ph]} \BibitemShut {NoStop}%
\bibitem [{\citenamefont {Kiczynski}\ \emph {et~al.}(2022)\citenamefont
  {Kiczynski}, \citenamefont {Gorman}, \citenamefont {Geng}, \citenamefont
  {Donnelly}, \citenamefont {Chung}, \citenamefont {He}, \citenamefont
  {Keizer},\ and\ \citenamefont {Simmons}}]{kiczynski2022engineering}%
  \BibitemOpen
  \bibfield  {author} {\bibinfo {author} {\bibfnamefont {M.}~\bibnamefont
  {Kiczynski}}, \bibinfo {author} {\bibfnamefont {S.~K.}\ \bibnamefont
  {Gorman}}, \bibinfo {author} {\bibfnamefont {H.}~\bibnamefont {Geng}},
  \bibinfo {author} {\bibfnamefont {M.~B.}\ \bibnamefont {Donnelly}}, \bibinfo
  {author} {\bibfnamefont {Y.}~\bibnamefont {Chung}}, \bibinfo {author}
  {\bibfnamefont {Y.}~\bibnamefont {He}}, \bibinfo {author} {\bibfnamefont
  {J.~G.}\ \bibnamefont {Keizer}},\ and\ \bibinfo {author} {\bibfnamefont
  {M.~Y.}\ \bibnamefont {Simmons}},\ }\bibfield  {title} {\bibinfo {title}
  {Engineering topological states in atom-based semiconductor quantum dots},\
  }\href {https://doi.org/10.1038/s41586-022-04706-0} {\bibfield  {journal}
  {\bibinfo  {journal} {Nature}\ }\textbf {\bibinfo {volume} {606}},\ \bibinfo
  {pages} {694} (\bibinfo {year} {2022})}\BibitemShut {NoStop}%
\bibitem [{\citenamefont {Zhao}\ \emph {et~al.}(2019)\citenamefont {Zhao},
  \citenamefont {Tanttu}, \citenamefont {Tan}, \citenamefont {Hensen},
  \citenamefont {Chan}, \citenamefont {Hwang}, \citenamefont {Leon},
  \citenamefont {Yang}, \citenamefont {Gilbert}, \citenamefont {Hudson},
  \citenamefont {Itoh}, \citenamefont {Kiselev}, \citenamefont {Ladd},
  \citenamefont {Morello}, \citenamefont {Laucht},\ and\ \citenamefont
  {Dzurak}}]{zhao2019singlespin}%
  \BibitemOpen
  \bibfield  {author} {\bibinfo {author} {\bibfnamefont {R.}~\bibnamefont
  {Zhao}}, \bibinfo {author} {\bibfnamefont {T.}~\bibnamefont {Tanttu}},
  \bibinfo {author} {\bibfnamefont {K.~Y.}\ \bibnamefont {Tan}}, \bibinfo
  {author} {\bibfnamefont {B.}~\bibnamefont {Hensen}}, \bibinfo {author}
  {\bibfnamefont {K.~W.}\ \bibnamefont {Chan}}, \bibinfo {author}
  {\bibfnamefont {J.~C.~C.}\ \bibnamefont {Hwang}}, \bibinfo {author}
  {\bibfnamefont {R.~C.~C.}\ \bibnamefont {Leon}}, \bibinfo {author}
  {\bibfnamefont {C.~H.}\ \bibnamefont {Yang}}, \bibinfo {author}
  {\bibfnamefont {W.}~\bibnamefont {Gilbert}}, \bibinfo {author} {\bibfnamefont
  {F.~E.}\ \bibnamefont {Hudson}}, \bibinfo {author} {\bibfnamefont {K.~M.}\
  \bibnamefont {Itoh}}, \bibinfo {author} {\bibfnamefont {A.~A.}\ \bibnamefont
  {Kiselev}}, \bibinfo {author} {\bibfnamefont {T.~D.}\ \bibnamefont {Ladd}},
  \bibinfo {author} {\bibfnamefont {A.}~\bibnamefont {Morello}}, \bibinfo
  {author} {\bibfnamefont {A.}~\bibnamefont {Laucht}},\ and\ \bibinfo {author}
  {\bibfnamefont {A.~S.}\ \bibnamefont {Dzurak}},\ }\bibfield  {title}
  {\bibinfo {title} {Single-spin qubits in isotopically enriched silicon at low
  magnetic field},\ }\href {https://doi.org/10.1038/s41467-019-13416-7}
  {\bibfield  {journal} {\bibinfo  {journal} {Nature Communications}\ }\textbf
  {\bibinfo {volume} {10}},\ \bibinfo {pages} {5500} (\bibinfo {year}
  {2019})}\BibitemShut {NoStop}%
\bibitem [{\citenamefont {Mohiyaddin}\ \emph {et~al.}(2013)\citenamefont
  {Mohiyaddin}, \citenamefont {Rahman}, \citenamefont {Kalra}, \citenamefont
  {Klimeck}, \citenamefont {Hollenberg}, \citenamefont {Pla}, \citenamefont
  {Dzurak},\ and\ \citenamefont {Morello}}]{mohiyaddin2013noninvasive}%
  \BibitemOpen
  \bibfield  {author} {\bibinfo {author} {\bibfnamefont {F.~A.}\ \bibnamefont
  {Mohiyaddin}}, \bibinfo {author} {\bibfnamefont {R.}~\bibnamefont {Rahman}},
  \bibinfo {author} {\bibfnamefont {R.}~\bibnamefont {Kalra}}, \bibinfo
  {author} {\bibfnamefont {G.}~\bibnamefont {Klimeck}}, \bibinfo {author}
  {\bibfnamefont {L.~C.~L.}\ \bibnamefont {Hollenberg}}, \bibinfo {author}
  {\bibfnamefont {J.~J.}\ \bibnamefont {Pla}}, \bibinfo {author} {\bibfnamefont
  {A.~S.}\ \bibnamefont {Dzurak}},\ and\ \bibinfo {author} {\bibfnamefont
  {A.}~\bibnamefont {Morello}},\ }\bibfield  {title} {\bibinfo {title}
  {Noninvasive spatial metrology of single-atom devices},\ }\href
  {https://doi.org/10.1021/nl303863s} {\bibfield  {journal} {\bibinfo
  {journal} {Nano Letters}\ }\textbf {\bibinfo {volume} {13}},\ \bibinfo
  {pages} {1903} (\bibinfo {year} {2013})}\BibitemShut {NoStop}%
\bibitem [{Note1()}]{Note1}%
  \BibitemOpen
  \bibinfo {note} {Unfortunately, the gate stack used in this device (palladium
  gate electrodes with atomic-layer deposited aluminum oxide) results in much
  higher charge noise than the previously used aluminum gate stack~\cite
  {vahapoglu2021coherent}, and renders this measurement noisier than previous
  measurements.}\BibitemShut {Stop}%
\bibitem [{\citenamefont {Leon}\ \emph {et~al.}(2020)\citenamefont {Leon},
  \citenamefont {Yang}, \citenamefont {Hwang}, \citenamefont {Lemyre},
  \citenamefont {Tanttu}, \citenamefont {Huang}, \citenamefont {Chan},
  \citenamefont {Tan}, \citenamefont {Hudson}, \citenamefont {Itoh},
  \citenamefont {Morello}, \citenamefont {Laucht}, \citenamefont
  {{Pioro-Ladri{\`e}re}}, \citenamefont {Saraiva},\ and\ \citenamefont
  {Dzurak}}]{leon2020coherent}%
  \BibitemOpen
  \bibfield  {author} {\bibinfo {author} {\bibfnamefont {R.~C.~C.}\
  \bibnamefont {Leon}}, \bibinfo {author} {\bibfnamefont {C.~H.}\ \bibnamefont
  {Yang}}, \bibinfo {author} {\bibfnamefont {J.~C.~C.}\ \bibnamefont {Hwang}},
  \bibinfo {author} {\bibfnamefont {J.~C.}\ \bibnamefont {Lemyre}}, \bibinfo
  {author} {\bibfnamefont {T.}~\bibnamefont {Tanttu}}, \bibinfo {author}
  {\bibfnamefont {W.}~\bibnamefont {Huang}}, \bibinfo {author} {\bibfnamefont
  {K.~W.}\ \bibnamefont {Chan}}, \bibinfo {author} {\bibfnamefont {K.~Y.}\
  \bibnamefont {Tan}}, \bibinfo {author} {\bibfnamefont {F.~E.}\ \bibnamefont
  {Hudson}}, \bibinfo {author} {\bibfnamefont {K.~M.}\ \bibnamefont {Itoh}},
  \bibinfo {author} {\bibfnamefont {A.}~\bibnamefont {Morello}}, \bibinfo
  {author} {\bibfnamefont {A.}~\bibnamefont {Laucht}}, \bibinfo {author}
  {\bibfnamefont {M.}~\bibnamefont {{Pioro-Ladri{\`e}re}}}, \bibinfo {author}
  {\bibfnamefont {A.}~\bibnamefont {Saraiva}},\ and\ \bibinfo {author}
  {\bibfnamefont {A.~S.}\ \bibnamefont {Dzurak}},\ }\bibfield  {title}
  {\bibinfo {title} {Coherent spin control of s-, p-, d- and f-electrons in a
  silicon quantum dot},\ }\href {https://doi.org/10.1038/s41467-019-14053-w}
  {\bibfield  {journal} {\bibinfo  {journal} {Nature Communications}\ }\textbf
  {\bibinfo {volume} {11}},\ \bibinfo {pages} {797} (\bibinfo {year}
  {2020})}\BibitemShut {NoStop}%
\bibitem [{\citenamefont {Anderson}\ \emph {et~al.}(2022)\citenamefont
  {Anderson}, \citenamefont {Gyure}, \citenamefont {Quinn}, \citenamefont
  {Pan}, \citenamefont {Ross},\ and\ \citenamefont
  {Kiselev}}]{anderson2022highprecision}%
  \BibitemOpen
  \bibfield  {author} {\bibinfo {author} {\bibfnamefont {C.~R.}\ \bibnamefont
  {Anderson}}, \bibinfo {author} {\bibfnamefont {M.~F.}\ \bibnamefont {Gyure}},
  \bibinfo {author} {\bibfnamefont {S.}~\bibnamefont {Quinn}}, \bibinfo
  {author} {\bibfnamefont {A.}~\bibnamefont {Pan}}, \bibinfo {author}
  {\bibfnamefont {R.~S.}\ \bibnamefont {Ross}},\ and\ \bibinfo {author}
  {\bibfnamefont {A.~A.}\ \bibnamefont {Kiselev}},\ }\bibfield  {title}
  {\bibinfo {title} {High-precision real-space simulation of electrostatically
  confined few-electron states},\ }\href {https://doi.org/10.1063/5.0089350}
  {\bibfield  {journal} {\bibinfo  {journal} {AIP Advances}\ }\textbf {\bibinfo
  {volume} {12}},\ \bibinfo {pages} {065123} (\bibinfo {year}
  {2022})}\BibitemShut {NoStop}%
\bibitem [{\citenamefont {Ercan}\ \emph {et~al.}(2021)\citenamefont {Ercan},
  \citenamefont {Coppersmith},\ and\ \citenamefont
  {Friesen}}]{ercan2021strong}%
  \BibitemOpen
  \bibfield  {author} {\bibinfo {author} {\bibfnamefont {H.~E.}\ \bibnamefont
  {Ercan}}, \bibinfo {author} {\bibfnamefont {S.~N.}\ \bibnamefont
  {Coppersmith}},\ and\ \bibinfo {author} {\bibfnamefont {M.}~\bibnamefont
  {Friesen}},\ }\bibfield  {title} {\bibinfo {title} {Strong electron-electron
  interactions in si/sige quantum dots},\ }\href
  {https://doi.org/10.1103/PhysRevB.104.235302} {\bibfield  {journal} {\bibinfo
   {journal} {Physical Review B}\ }\textbf {\bibinfo {volume} {104}},\ \bibinfo
  {pages} {235302} (\bibinfo {year} {2021})}\BibitemShut {NoStop}%
\bibitem [{\citenamefont {Pratt}(1956)}]{pratt1956unrestricted}%
  \BibitemOpen
  \bibfield  {author} {\bibinfo {author} {\bibfnamefont {G.~W.}\ \bibnamefont
  {Pratt}},\ }\bibfield  {title} {\bibinfo {title} {Unrestricted {Hartree-Fock}
  method},\ }\href {https://doi.org/10.1103/PhysRev.102.1303} {\bibfield
  {journal} {\bibinfo  {journal} {Physical Review}\ }\textbf {\bibinfo {volume}
  {102}},\ \bibinfo {pages} {1303} (\bibinfo {year} {1956})}\BibitemShut
  {NoStop}%
\bibitem [{\citenamefont {Yang}\ \emph {et~al.}(2011)\citenamefont {Yang},
  \citenamefont {Lim}, \citenamefont {Zwanenburg},\ and\ \citenamefont
  {Dzurak}}]{Yang2011dynamically}%
  \BibitemOpen
  \bibfield  {author} {\bibinfo {author} {\bibfnamefont {C.~H.}\ \bibnamefont
  {Yang}}, \bibinfo {author} {\bibfnamefont {W.~H.}\ \bibnamefont {Lim}},
  \bibinfo {author} {\bibfnamefont {F.~A.}\ \bibnamefont {Zwanenburg}},\ and\
  \bibinfo {author} {\bibfnamefont {A.~S.}\ \bibnamefont {Dzurak}},\ }\bibfield
   {title} {\bibinfo {title} {Dynamically controlled charge sensing of a
  few-electron silicon quantum dot},\ }\href
  {https://doi.org/10.1063/1.3654496} {\bibfield  {journal} {\bibinfo
  {journal} {AIP Advances}\ }\textbf {\bibinfo {volume} {1}},\ \bibinfo {pages}
  {042111} (\bibinfo {year} {2011})}\BibitemShut {NoStop}%
\bibitem [{\citenamefont {Kindt}\ and\ \citenamefont
  {Van~Zeijl}(1998)}]{kindt1998modelling}%
  \BibitemOpen
  \bibfield  {author} {\bibinfo {author} {\bibfnamefont {W.}~\bibnamefont
  {Kindt}}\ and\ \bibinfo {author} {\bibfnamefont {H.}~\bibnamefont
  {Van~Zeijl}},\ }\bibfield  {title} {\bibinfo {title} {Modelling and
  fabrication of {Geiger} mode avalanche photodiodes},\ }\href
  {https://doi.org/10.1109/23.682621} {\bibfield  {journal} {\bibinfo
  {journal} {IEEE Transactions on Nuclear Science}\ }\textbf {\bibinfo {volume}
  {45}},\ \bibinfo {pages} {715} (\bibinfo {year} {1998})}\BibitemShut
  {NoStop}%
\bibitem [{\citenamefont {Botkin}\ and\ \citenamefont
  {Turova}(2004)}]{botkin2004mathematical}%
  \BibitemOpen
  \bibfield  {author} {\bibinfo {author} {\bibfnamefont {N.~D.}\ \bibnamefont
  {Botkin}}\ and\ \bibinfo {author} {\bibfnamefont {V.~L.}\ \bibnamefont
  {Turova}},\ }\bibfield  {title} {\bibinfo {title} {Mathematical models of a
  biosensor},\ }\href {https://doi.org/10.1016/j.apm.2003.10.012} {\bibfield
  {journal} {\bibinfo  {journal} {Applied Mathematical Modelling}\ }\textbf
  {\bibinfo {volume} {28}},\ \bibinfo {pages} {573} (\bibinfo {year}
  {2004})}\BibitemShut {NoStop}%
\bibitem [{\citenamefont {Vahapoglu}\ \emph {et~al.}(2021)\citenamefont
  {Vahapoglu}, \citenamefont {{Slack-Smith}}, \citenamefont {Leon},
  \citenamefont {Lim}, \citenamefont {Hudson}, \citenamefont {Day},
  \citenamefont {Cifuentes}, \citenamefont {Tanttu}, \citenamefont {Yang},
  \citenamefont {Saraiva}, \citenamefont {Abrosimov}, \citenamefont {Pohl},
  \citenamefont {Thewalt}, \citenamefont {Laucht}, \citenamefont {Dzurak},\
  and\ \citenamefont {Pla}}]{vahapoglu2021coherent}%
  \BibitemOpen
  \bibfield  {author} {\bibinfo {author} {\bibfnamefont {E.}~\bibnamefont
  {Vahapoglu}}, \bibinfo {author} {\bibfnamefont {J.~P.}\ \bibnamefont
  {{Slack-Smith}}}, \bibinfo {author} {\bibfnamefont {R.~C.~C.}\ \bibnamefont
  {Leon}}, \bibinfo {author} {\bibfnamefont {W.~H.}\ \bibnamefont {Lim}},
  \bibinfo {author} {\bibfnamefont {F.~E.}\ \bibnamefont {Hudson}}, \bibinfo
  {author} {\bibfnamefont {T.}~\bibnamefont {Day}}, \bibinfo {author}
  {\bibfnamefont {J.~D.}\ \bibnamefont {Cifuentes}}, \bibinfo {author}
  {\bibfnamefont {T.}~\bibnamefont {Tanttu}}, \bibinfo {author} {\bibfnamefont
  {C.~H.}\ \bibnamefont {Yang}}, \bibinfo {author} {\bibfnamefont
  {A.}~\bibnamefont {Saraiva}}, \bibinfo {author} {\bibfnamefont {N.~V.}\
  \bibnamefont {Abrosimov}}, \bibinfo {author} {\bibfnamefont {H.-J.}\
  \bibnamefont {Pohl}}, \bibinfo {author} {\bibfnamefont {M.~L.~W.}\
  \bibnamefont {Thewalt}}, \bibinfo {author} {\bibfnamefont {A.}~\bibnamefont
  {Laucht}}, \bibinfo {author} {\bibfnamefont {A.~S.}\ \bibnamefont {Dzurak}},\
  and\ \bibinfo {author} {\bibfnamefont {J.~J.}\ \bibnamefont {Pla}},\ }\href
  {https://doi.org/10.48550/arXiv.2107.14622} {\bibinfo {title} {Coherent
  control of electron spin qubits in silicon using a global field}} (\bibinfo
  {year} {2021}),\ \Eprint {https://arxiv.org/abs/2107.14622} {arXiv:2107.14622
  [cond-mat]} \BibitemShut {NoStop}%
\end{thebibliography}%

% Biography Section
%\include{bio/biography}

\end{document}